\documentclass[universe,article,accept,pdftex,moreauthors]{Definitions/mdpi}
\firstpage{1}
\pubvolume{12}
\issuenum{3}
\articlenumber{77}
\pubyear{2026}
\copyrightyear{2026}
\externaleditor{Piergiorgio Casella} 
\datereceived{10 December 2025}
\daterevised{1 March 2026} 
\dateaccepted{4 March 2026}
\datepublished{10 March 2026}
\hreflink{https://doi.org/10.3390/universe12030077} 
\doinum{10.3390/universe12030077}
\pdfoutput=1 

\Title{Magnetohydrodynamic Simulations of Transonic Accretion Flows}

\TitleCitation{Magnetohydrodynamic Simulations of Transonic Accretion Flows}

\Author{Raj Kishor Joshi $^{1}$\orcidA{}*, Antonios Tsokaros $^{2,3,4}$\orcidB{}, Sanjit Debnath $^{5}$\orcidC{}, 
        Indranil Chattopadhyay $^{5}$\orcidD{}, and Ramiz Aktar $^{6}$\orcidE{}  
}
\AuthorNames{Raj Kishor Joshi, Antonios Tsokaros, Sanjit Debnath, Indranil Chattopadhyay and Ramiz Aktar}

\isAPAStyle{%
       \AuthorCitation{Joshi, R.K., Tsokaros, A., Debnath, S., Chattopadhyay, I., \& Aktar, R.}
         }{%
        \isChicagoStyle{%
        \AuthorCitation{Joshi, Raj Kishor, Antonios Tsokaros, Sanjit Debnath, Indranil Chattopadhyay, and Ramiz Aktar.}
        }{
        \AuthorCitation{Joshi, R.K.; Tsokaros, A.; Debnath, S.; Chattopadhyay, I.; Aktar, R.}
        }
}

\address{%
$^{1}$ \quad Nicolaus Copernicus Astronomical Center of the Polish Academy of Sciences, Bartycka 18, PL-00-716 Warsaw, Poland \\
$^{2}$ \quad Department of Physics, University of Illinois at Urbana-Champaign, Urbana, IL 61801, USA; tsokaros@illinois.edu\\
$^{3}$ \quad National Center for Supercomputing Applications, University of Illinois at Urbana-Champaign, Urbana, IL 61801, USA\\
$^{4}$ \quad Research Center for Astronomy and Applied Mathematics, Academy of Athens, 11527 Athens, Greece\\
$^{5}$ \quad Aryabhatta Research Institute of Observational Sciences (ARIES), Manora Peak, Nainital 263001, India; sdebnath@aries.res.in (S.D.); indra@aries.res.in (I.C.)\\
$^{6}$ \quad Institute of Astronomy and Astrophysics, Academia Sinica, Taipei 10617, Taiwan; mraktar@asiaa.sinica.edu.tw
}

\corres{Correspondence: rjoshi@camk.edu.pl}

\abstract{Theoretical studies of transonic accretion onto black holes reveal a wide range of possible solutions, broadly classified into smooth flows and flows featuring shocks. Accretion solutions that involve the formation of shocks are particularly intriguing, as they are expected to naturally produce observable variability features. However, despite their theoretical significance, time-dependent studies exploring the stability and evolution of such shocked solutions remain relatively scarce. To address this gap, we perform simulations of transonic accretion flows around a black hole in an ideal magnetohydrodynamic framework. Our simulations are initialized using boundary conditions derived from semi-analytical hydrodynamical models, allowing us to explore the stability of these flows under varying magnetic field strengths. Our results indicate that mildly magnetized flows in a uniform vertical magnetic field alter the accretion dynamics through magnetic pressure, with the resulting force imbalance driving oscillations in the shock front. Variations in the emitted luminosity arising from shock oscillations appear as quasi-periodic oscillations (QPOs), a characteristic feature commonly observed in accreting black holes. We find that the QPO frequency is determined by the radial position of the shock front: oscillations occurring closer to the black hole produce frequencies of tens of hertz, whereas shocks located farther out yield sub-hertz frequencies.}


\usepackage{graphicx}	
\usepackage{amsmath}	

\usepackage{comment}

\begin{document}


\section{Introduction}
Accretion onto compact objects such as black holes is one of the most fundamental processes governing high-energy astrophysical phenomena. The infall of matter into the gravitational potential well of a black hole releases enormous amounts of energy, which is often observed as intense radiation from accreting systems such as black hole X-ray binaries (BHXRBs) and active galactic nuclei (AGN) \citep{1973A&A....24..337S,1972A&A....21....1P}.
BHXRBs display pronounced spectral and temporal variability across different energy bands. Their observational properties strongly depend on the accretion state, with transitions between spectral states accompanied by significant changes in timing and radiative characteristics. In particular, BHXRBs can exhibit quasi-periodic oscillations (QPOs) in the power-law component of their X-ray spectra, with both phenomena being closely linked to the underlying accretion dynamics and spectral state of the system \citep[see, e.g.,][]{2004MNRAS.355.1105F,2019NewAR..8501524I,2024MNRAS.531.1149N}.
Understanding the dynamics and energetics of such accretion flows is therefore crucial for interpreting astrophysical observations of black hole candidates.


Depending on the angular momentum of the accreting matter, two idealized limiting solutions can be identified. In the absence of angular momentum, the flow is spherically symmetric, leading to the classical Bondi accretion solution \citep{1952MNRAS.112..195B}, where matter falls radially onto the compact object. In contrast, when the angular momentum is sufficiently high, the accreting material can settle into a geometrically thin, nearly Keplerian disk, in which radial motion is negligible and the flow is primarily rotational \citep{1973A&A....24..337S}. While these two extremes provide useful conceptual frameworks, realistic accretion flows around black holes are neither purely radial nor perfectly Keplerian.
In a more general scenario, rotating accretion flows with significant radial advection introduce a much richer dynamical structure through the interplay between centrifugal forces and advection velocities. 
Such transonic solutions are crucial for understanding the global structure of accretion flows, as they determine how matter can smoothly transition from subsonic to supersonic speeds while satisfying the inner boundary condition at the event horizon. The resulting flow topologies are diverse, ranging from single-sonic-point solutions similar to spherical Bondi accretion to accretion flows that can exhibit multiple sonic points \citep{1987PASJ...39..309F,2014MNRAS.443.3444K,2011IJMPD..20.1597C}, which play a key role in regulating the thermodynamic and radiative properties of the accreting plasma.
This transonic advective disk can have standing shocks, where a portion of the bulk kinetic energy of the flow is converted into thermal energy \citep{2008ApJ...677L..93B, 2009ApJ...702..649D, 2011IJMPD..20.1597C, 2013MNRAS.430..386K, 2014MNRAS.443.3444K}. The resulting hot post-shock region can serve as a Comptonizing corona that upscatters soft photons, thereby influencing the observed emission spectrum of the accretion system \citep{1995ApJ...455..623C, 2020A&A...642A.209S, 2023MNRAS.522.3735S}.

However, several models have been proposed to explain QPOs in the BHXRBs, including the relativistic precession model (RPM) \citep{1998ApJ...492L..59S, 1999ApJ...524L..63S}, the precessing inner flow model \citep{2009MNRAS.397L.101I}, the accretion--ejection instability \citep{1999A&A...349.1003T}, the propagating oscillatory shock model \citep{1996ApJ...457..805M, 2008A&A...489L..41C}, and the magneto-rotational instability (MRI) turbulence model \citep{1999ApJ...521..650B}.
Focusing on the proposed propagating oscillatory shock model, a series of inviscid hydrodynamic simulations have been performed to study the formation and evolution of shocked, transonic accretion disks \citep{1994ApJ...425..161M, 1996ApJ...457..805M, 1996ApJ...470..460M, 1995ApJ...452..364R, 2010MNRAS.403..516G, 2012MNRAS.425.2413O, 2012ApJ...758..114G, 2014MNRAS.437.1329G, 2017MNRAS.472.4327S, 2017MNRAS.472..542K, 2019MNRAS.482.3636K, 2020ApJ...904...21P, 2022MNRAS.514.5074O, 2023A&A...678A.141O, 2023ApJ...946L..42K, 2025ApJ...990...12M, 2023MNRAS.519.4550G}. Because shock formation is closely tied to the angular momentum distribution within the disk, the inclusion of viscosity, which governs angular momentum transport, is expected to influence both the existence and stability of the shock significantly. Indeed, several studies have shown that viscosity can induce oscillations in shock position \citep{1998MNRAS.299..799L, 2011ApJ...728..142L, 2012MNRAS.421..666G, 2013MNRAS.430.2836G, 2014MNRAS.442..251D, 2015MNRAS.448.3221G, 2015MNRAS.453..147O, 2016ApJ...831...33L, 2024MNRAS.528.3964D}. 
The characteristic frequencies of these oscillations are found to closely resemble the quasi-periodic oscillations (QPOs) observed in BHXRBs, suggesting that the underlying shock dynamics may provide a natural explanation for these observed luminosity modulations. Recent investigations by \citet{2024MNRAS.528.3964D,2025ApJ...994...48D} have further demonstrated that the temporal variations in luminosity arising from viscous, cooling-modulated shock oscillations can reproduce QPOs in the sub-Hertz to few-Hertz range for microquasars hosting black holes of tens of solar masses.

Magnetic fields in accretion disks play a crucial role in driving angular momentum transport and energy dissipation through the magneto-rotational instability (MRI) \citep{1991ApJ...376..214B,1992ApJ...400..595H}. The inclusion of magnetic fields introduces additional physical processes, such as magnetic pressure, tension, and reconnection, that can substantially alter the global accretion dynamics.
Over the past few decades, numerous general relativistic magnetohydrodynamic (GRMHD) codes have been developed and applied to a wide range of astrophysical problems. They have been extensively used to investigate accretion processes and jet formation, among other phenomena \citep{1984ApJS...55..211H,2003ApJ...589..444G,2009ApJ...703L.142D,2016A&A...586A..38M,2011MNRAS.418L..79T,2014MNRAS.441.3177M,2012ApJS..201....9F,2007ApJ...668..417F,2014MNRAS.439..503S,2015MNRAS.447...49S,2018MNRAS.474L..81L,2025ApJ...980..203D}. GRMHD simulations of tilted black hole accretion disks have revealed a rich interplay between relativistic frame dragging, magnetic turbulence, and nonlinear disk dynamics: differential nodal precession driven by Lense–Thirring precession causes the inner regions of a tilted disk around a Kerr black hole to warp, and strong external torques can overwhelm internal magnetic stresses and lead to disk tearing, where the flow breaks into discrete, independently precessing rings that enhance dissipation and can drive variability in accretion rate and jet orientation; these dynamics directly influence jet behavior powered through the Blandford–Znajek process \citep{2025PhRvD.112f3013C,2025ApJ...995..112J}. Subgrid models have also been used to accurately capture MRI-driven turbulence and to study intrinsic dynamo processes \citep{2025arXiv251203443Z}.
For a comprehensive overview of the modern GRMHD view of accretion disk simulations, see also \citep{2019ApJS..243...26P}. Recent GRMHD studies \citep{2022ApJ...936L...5L, 2023ApJ...946L..42K, 2024ApJ...964...79L,2025ApJ...978..148G,2025ApJ...985..135C} have also explored the different types of low-angular-momentum accretion flows, particularly showing an outward shock structure in the disk and formation of jets.
Despite the development of GRMHD code, systematic investigations of transonic accretion flows with shocks in the magnetohydrodynamic (MHD) regime remain relatively scarce.
Recent general relativistic magnetohydrodynamic (GRMHD) simulations by \citet{2025ApJ...990...12M} and \citet{2025arXiv250723187D} have shown that magnetized flows around rotating black holes can sustain standing shocks. Shock formation in accretion flows onto binary black hole systems has also been reported by \citet{2025arXiv250916796K}. Although the number of such studies is still limited, they collectively demonstrate that shocks can also form in magnetized accretion flows. However, the role of magnetic field strength and topology in regulating the onset, stability, and oscillatory behavior of these shocks remains poorly constrained. Addressing these issues is crucial for developing a self-consistent framework that links the microphysics of magnetized accretion flows to the macroscopic variability observed in black hole systems. To bridge this gap, in this work, we present results from magnetohydrodynamic (MHD) simulations of transonic accretion flows around a black hole, with an emphasis on the stability and temporal evolution of shock solutions. The simulations are initialized using boundary conditions derived from semi-analytical hydrodynamical solutions, which generate steady-state shocks. Our objective is to investigate the time-dependent behavior of such solutions by analyzing the resulting flow morphology, evolution, and stability of shocks in the flow, and corresponding luminosity variations.

The remainder of the paper is organized as follows. Section \ref{sec:setup} describes the theoretical framework and numerical setup used in our simulations. Section \ref{sec:results} presents the main results, focusing on the role of magnetic field strength in determining flow stability and shock dynamics. Finally, Section \ref{sec:concl} summarizes our conclusions and outlines potential directions for future work.

\section{Methods}
\label{sec:setup}
Although analytical solutions for transonic accretion exist in the fully general relativistic regime, we restrict our analysis to the classical Newtonian MHD limit. This non-relativistic framework is chosen to isolate and examine the specific physical processes of interest. In particular, we focus on the role of magnetic fields in modifying shock dynamics and driving oscillatory behavior in transonic accretion flows, as well as on evaluating the extent to which the shock--oscillation model can explain QPOs over a broad range of frequencies. We note that the Newtonian MHD treatment adopted here does not capture relativistic effects that become important near the inner boundary, including frame dragging and relativistic corrections to the flow energetics. In addition, jet launching is not addressed in this framework, as spin-driven energy extraction require a fully relativistic treatment.
We model the accretion flow using the publicly available code \texttt{PLUTO} \citep{pluto07} in the ideal MHD limit.
\texttt{PLUTO} uses the higher-order Godunov scheme to solve the equations

\begin{equation}
\frac{\partial \rho}{\partial t}+\nabla \cdot\left(\rho \mathbf{v}\right)=0\, ,  
\label{eq:continuity}
\end{equation}

\begin{equation}
\frac{\partial (\rho \mathbf{v})}{\partial t}+\nabla \cdot\left(\rho \mathbf{v\otimes  v-B\otimes B}\right)+\nabla p_\text{t}=\rho\nabla\Phi_\text{PW}\,  ,  
\label{eq:euler}
\end{equation}

\begin{equation}
\frac{\partial \mathbf{B} }{\partial t}+\nabla \times\left(\mathbf{v}\times \mathbf{B}\right)=0 \, ,
\label{eq:induction}
\end{equation}

\begin{equation}
\frac{\partial E}{\partial t}+\nabla \cdot\left[(E+p_\text{t})\mathbf{v}-(\mathbf{v\cdot B})\mathbf{B}\right]=\rho\mathbf{v}\cdot\nabla\Phi_\text{PW} \, ,
\label{eq:energy}
\end{equation}
here $\Phi_\text{PW}=-GM/(r-r_g)$ is the Paczy{\'n}sky-Wiita potential \citep{1980A&A....88...23P} and $r_g=2GM/c^2$ is the Schwarzschild radius. $p_\text{t} = p + p_\text{mag} $ is the
total pressure consisting of thermal ($p$), and magnetic ($p_\text{mag}=|\mathbf{B}|^2/2$)
pressure. The total energy density $E$ is the sum of the internal, kinetic and magnetic energy density given as   
\begin{equation}
E=e+\frac{1}{2}\rho|\mathbf{v}|^2+\frac{|\mathbf{B}|^2}{2} \, ,
\label{tot_e}    
\end{equation}
In addition to the above equations the magnetic field satisfies the solenoidal constraint,
\begin{equation}
\nabla \cdot\mathbf{B}=0.
\label{eq:divb}
\end{equation}
To close the system of equations, an additional relation is required that connects the internal energy density $e$ to the fluid variables $\rho$ and $p$. We adopt an ideal gas equation of state (EoS) for which the internal energy density is given as $e = p/(\gamma - 1)$. The accreting matter is cold at large distances from the black hole and becomes progressively hotter as it approaches the event horizon. Consequently, the adiabatic index should transition from a higher value ($\gamma = 5/3$) in the outer colder regions to a lower value ($\gamma = 4/3$) in the hotter inner regions. Accurately capturing this thermal evolution would require an EoS similar to the EoS given by \citet{m71,rc06,2009ApJ...694..492C} that self-consistently computes the adiabatic index from the local temperature. In the present study, such an equation of state was not incorporated into our simulation framework, and we therefore employed the fixed $\gamma$ equation of state provided in \texttt{PLUTO}, adopting $\gamma = 4/3$.
   

These equations are solved on a spherical grid ($r,\,\theta,\,\phi$) using the HLLC Riemann solver \citep{Harten_1983} with piecewise parabolic reconstruction \citep{cw84}.  
Time evolution is performed using a second-order Runge–Kutta (RK2) time-stepping method. 
To maintain the divergence-free condition of the magnetic field, we employ the hyperbolic divergence cleaning method of \citet{dedner02}. 

We adopt a unit system in which the unit of length is the Schwarzschild radius, $r_g = 2GM / c^2$, and the velocity is measured in units of the speed of light, $c$. 
Hence, the unit of time is $t_g = 2GM / c^3$, where $M$ is the mass of the black hole. In this unit system, the unit of specific angular momentum is $r_g c$.
Since our simulations do not include any dissipative processes, they are scale-free, and the length and time scales can be translated to those associated with either microquasars hosting stellar mass black holes or AGNs containing supermassive black holes, depending on the mass of the central object.

\subsection{Initial and Boundary Conditions}
The computational domain is initially filled with a tenuous, static atmosphere with a rest-mass density 
$\rho_{\rm atm} = 10^{-5}$ and gas pressure $p_{\rm atm} = 10^{-7}$. These parameters can be converted into physical units based on the specific value of mass accretion rate considered for the problem, as detailed in section \ref{sec:lum}. 
Matter is injected through the outer radial boundary; the injection parameters, the radial velocity $v$ 
and the dimensionless temperature $\Theta \equiv p / \rho$ are taken from semi-analytical 
steady-state transonic solutions (see \citealt{2025ApJ...994...48D} for details). The radial domain extends from $r=1.8$ to $r=200$. 
The inflow boundary condition is applied over an angular region $|\theta - \pi / 2| \leq \pi/6$ at the outer radius, 
while the remainder of that boundary uses outflow (zero-gradient) boundary conditions. For subsonic and sub-Alfv\'enic flows, ideal MHD permits upstream propagation of Alfv\'en and magnetosonic waves. In such cases, the number of boundary conditions that may be prescribed is limited by the number of incoming characteristics of the hyperbolic MHD system. Imposing all primitive variables at the injection boundary would therefore over-constrain the system and can lead to artificial wave reflections.
In our simulations, we impose inflow boundary conditions only on the hydrodynamic variables ($\rho$, $p$, $v_r$, and $v_\phi$), while $v_\theta$ and all magnetic field components are treated using continuous (zero-gradient) conditions. This allows the outgoing MHD characteristics to be determined self-consistently by the interior solution and avoids over-specification of the magnetic degrees of freedom.
Furthermore, since $\beta >> 1$ near the outer boundary, the flow is gas-pressure-dominated and the fast magnetosonic speed is close to the sound speed. The boundary treatment, therefore, closely approximates the characteristic structure of the corresponding hydrodynamic problem. We have verified that no spurious reflections or boundary-driven artifacts are present in the solution.
For 2D axisymmetric runs, we impose axisymmetric boundary conditions along the polar axis in $\theta$. 
For 3D runs, we use the \texttt{polaraxis} boundary condition available in \texttt{PLUTO} along the $\theta$ direction to avoid pole singularity. This boundary condition means copying the values to the ghost zones from the active zones after a rotation of $\pi$ around the pole. The velocity components (except the parallel one) reverse sign. We use periodic boundary conditions along $\phi$.

\vspace{0.5em}
\noindent
We used a grid resolution of $N_r \times N_\theta = 200 \times 296$ for 2D axisymmetric simulations and 
$N_r \times N_\theta \times N_\phi = 100 \times 296 \times 120$ for the fully 3D run.
The radial grid is logarithmically spaced to achieve higher resolution near the black hole.
Additionally, the $\theta$-grid has high resolution near the equator. The region $\pi/4-3\pi/4$ is resolved by 200 cells.
The choice of resolution and grid structure was made to optimize the ability of the code to capture shocks efficiently
without incurring excessive computational cost. The resolution in the radial direction was tuned by matching the results of the 1D simulation with the analytical solution. For multidimensional simulations, the resolution along the $\theta$ direction also becomes important to capture a steady-shock transition. We performed a series of numerical simulations for the unmagnetized flow to determine that a resolution of 100 cells in the region $\pi/2 \pm \pi/4$ is sufficient to capture a steady shock.
A recent study by \citet{2025arXiv250723187D} has shown that, in the weak-field limit, different initial magnetic field configurations, specifically inclined versus vertical fields, produce qualitatively similar density distributions, velocity streamlines, and Mach number profiles. Motivated by this result, we adopt a simpler configuration in the present study and prescribe a uniform vertical magnetic field at $t = 0$ of the form
\begin{equation}
B = B_0\,\hat{z},
\end{equation}
where $B_0$ is set to achieve the desired initial plasma beta in the whole domain, 
\begin{equation}
\beta_0 \equiv \frac{2p}{B_0^2} 
\end{equation}
Different runs use different initial $\beta$ values to explore the dependence of the flow on magnetic field strength.

In order to avoid very small time steps caused by the unbounded nature of the Alfvén velocity in the non-relativistic MHD regime, 
we impose a ceiling on the magnetization parameter, defined as $\sigma = B^2 / \rho$, with a threshold value of 100. 
We also set a floor value for the plasma $\beta$ parameter at $\beta = 0.01$. 
Any cells exceeding these thresholds are flagged, and their values are reset to the prescribed limits. 
The simulations are evolved for long durations (typically $t \gtrsim 5 \times 10^4\, t_g$, where $t_g \equiv 2GM / c^3$) 
to assess whether the system relaxes to a steady-state configuration or exhibits sustained time-dependent variability. 
The robustness of the numerical scheme and its ability to resolve shocks at the adopted resolution were verified 
by reproducing the corresponding one-dimensional solution, as presented in Appendix \ref{sec:app1}.

\vspace{0.5em}
\noindent
\vspace{0.5em}
\noindent

\section{Results}
\label{sec:results}

We begin by presenting the results of the purely hydrodynamic simulation, which serves as the fiducial reference case for this study. This run establishes the baseline morphology of the shocked accretion flow and provides a benchmark for assessing deviations from the underlying steady-state analytical solution, ensuring that any departures can be attributed to genuine physical effects rather than numerical artifacts. \\

We then analyze the magnetized simulations, which display clear morphological differences relative to the hydrodynamic case, demonstrating the influence of magnetic fields on the accretion structure and the resulting time-dependent behavior. 
Finally, we compute the bolometric luminosities from all simulations through a posteriori radiative estimates, showing that the temporal evolution of the shocks produces a characteristic imprint on the emitted radiation.

\subsection{Non-Magnetized Run}

 \begin{figure*}[h]
	\begin{center}       \includegraphics[width=\textwidth]{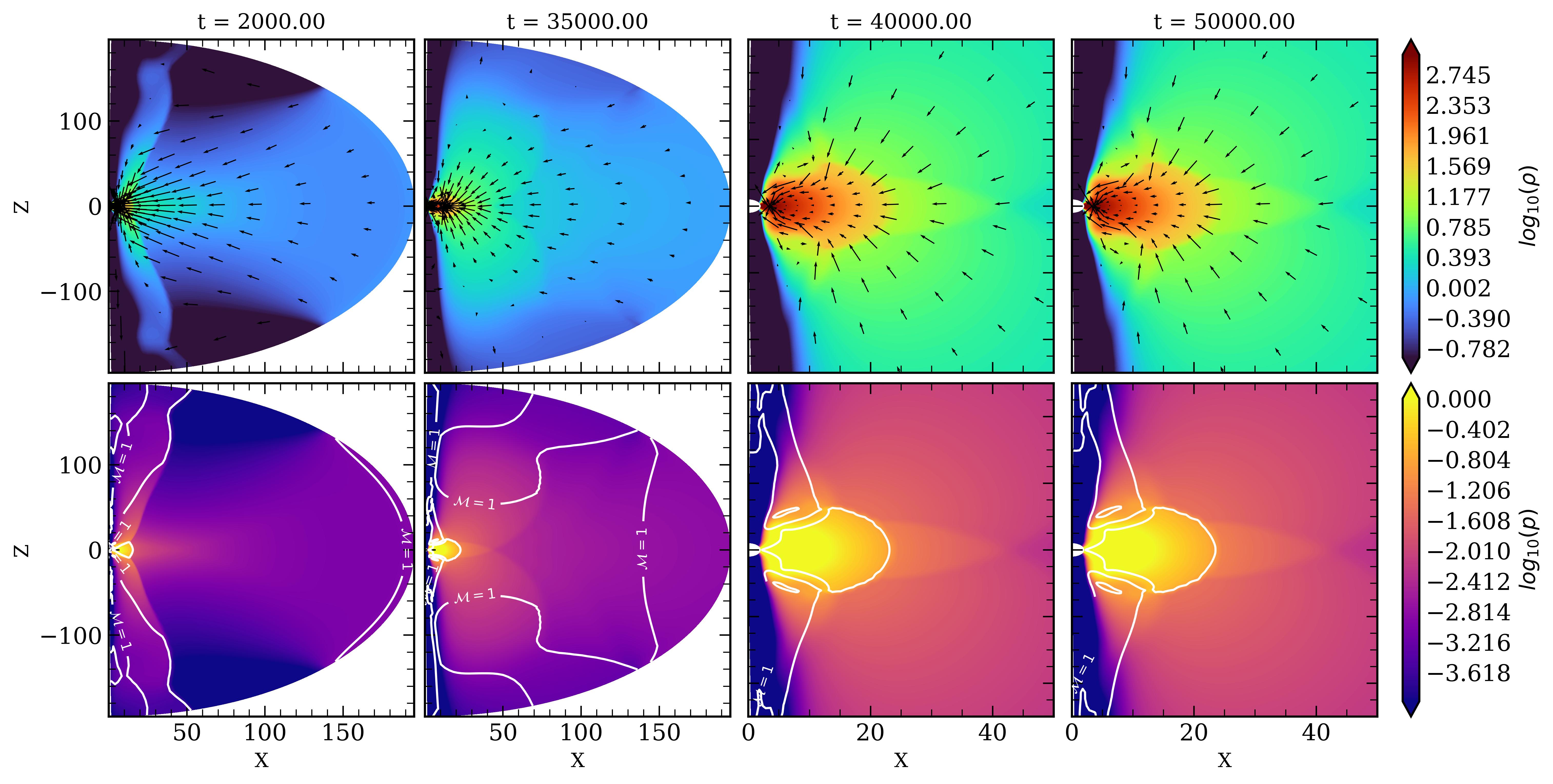}
        \caption{Top row: Density snapshots of the non-magnetized run \texttt{RUN0}. Arrows indicate the fluid velocity. Bottom row: Pressure snapshots with white contours indicating shock and sonic surfaces. Units of space and time are in terms of $r_g$ and $t_g$ correspondingly.} 
        \label{fig:hydro}
        \end{center}
\end{figure*}

The fiducial simulation \texttt{RUN0} is carried out under non-magnetized conditions. In this model, the accretion flow evolves solely under the influence of gravity and gas pressure. The solution is specified by a fixed angular momentum $\lambda = 1.75$ and a Bernoulli parameter $\mathcal{E} = 0.005$.
Our simulation strategy for transonic accretion flows differs from approaches that employ Fishbone--Moncrief tori \citep{1976ApJ...207..962F, 2024ApJ...972...18A,2024MNRAS.527.1745A} or thin-disk initial conditions \citep{1973blho.conf..343N,2020MNRAS.492.1855M}, where the computational domain is pre-filled with a prescribed disk profile. In contrast, for transonic flows, the domain is initially set to very low density and pressure floor values, and matter is continuously injected through the outer radial boundary. The injection parameters are taken directly from the analytical transonic solution, with $v_{\rm inj} = -0.0198$ and $\Theta_{\rm inj} = 0.00182$ at the outer boundary, which is at $r_{\rm inj} = 200$. The density at the outer boundary is set to $\rho=1$, and $v_\phi$ is calculated using $v_\phi=\lambda/r_{inj}$. \\

In the top row of Fig.~\ref{fig:hydro}, we show logarithmic density contours overlaid with velocity vectors to illustrate the flow dynamics. The injected material initially takes some time to propagate inward and reach the inner boundary, as seen in the first panel. Once the inflow establishes itself, matter begins to accumulate and gradually settles into a steady configuration. The bottom row shows the corresponding pressure distribution, with a contour indicating the surface where the radial Mach number $\mathcal{M} = v^r/c_s$ reaches unity, $c_s = \sqrt{\gamma \Theta}$ is the sound-speed. The chosen injection parameters impose a subsonic inflow at the outer boundary. As the gas moves inward under gravity, it accelerates and eventually becomes supersonic. At $t = 35{,}000\,t_g$, the contour $\mathcal{M}=1$ identifies the sonic surface at approximately $r \sim 140$. \\

By $t = 40{,}000\,t_g$, a sufficient amount of matter has accumulated to form a shock surface, with the shock located near $r \sim 25$ on the equatorial plane ($\theta = \pi/2$). The last two columns of the figure provide a zoomed-in view of the inner disk structure. A comparison between the snapshots at $t = 40{,}000\,t_g$ and $t = 50{,}000\,t_g$ demonstrates that the overall morphology approaches a quasi-steady state characterized by a well-defined standing shock. Although the one-dimensional analytical estimate predicts the shock location around $r \sim 10$, an exact match is not expected in multidimensional simulations, as additional degrees of freedom allow the Rankine--Hugoniot conditions \citep{1989ApJ...347..365C} to be satisfied at different radii. The hydrodynamic run also exhibits clear equatorial symmetry about the mid-plane, consistent with the expected behavior of an axisymmetric accretion flow in the absence of perturbations \citep{2023MNRAS.519.4550G}.

\subsection{Magnetized Simulations}

\begin{figure*}[h]
	\begin{center}       \includegraphics[width=\textwidth]{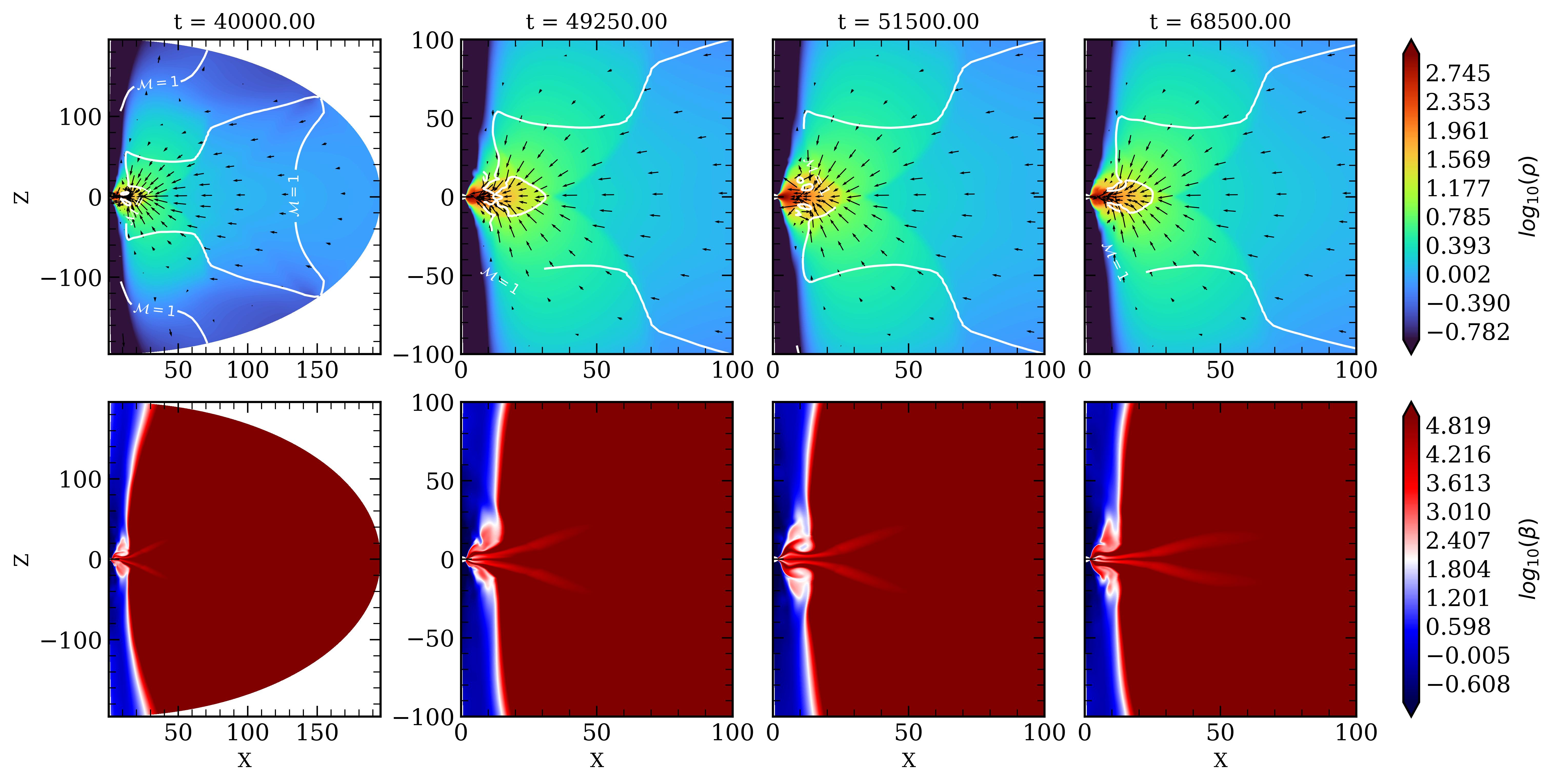}
        \caption{Density contours and distribution of plasma $\beta$ for model $\texttt{RUN1}$.}
        \label{fig:beta1e5}
        \end{center}
\end{figure*}

We now turn to the magnetized simulations, which explore how varying magnetic field strengths affect the accretion dynamics. We initialize the flows with three different plasma $\beta$ values: $\beta_0 = 10^5$, $5\times10^4$, and $10^3$. The corresponding models, labeled \texttt{RUN1}, \texttt{RUN2}, and \texttt{RUN3}, are identical in all aspects except for this choice of initial magnetization.
The maximum magnetization considered in this study is restricted by the specific region of parameter space relevant to our objectives, namely that yielding oscillatory or transient shock solutions, and by the unbounded nature of the Alfvén velocity in the Newtonian MHD framework at higher magnetization.
As the steady-shock solution is obtained by a perfect balance of forces in a pure hydrodynamic regime, a gradual departure from the steady state is expected if the magnetic field strength is increased.
We start with model \texttt{RUN1}; the density contours with overlaid velocity vectors are plotted in the top row of Fig. \ref{fig:beta1e5}, and plasma $\beta$ is plotted in the bottom row.
For MHD runs we plot the fast magnetosonic critical surface, identified numerically as the locus satisfying
\begin{equation}
M = 1,
\end{equation}
The fast magnetosonic Mach number is defined as
\begin{equation}
M = \frac{|v_r|}{c_f},
\end{equation}
where $v_r$ is the radial velocity component and the fast magnetosonic speed for radial propagation is obtained from the MHD dispersion relation,
\begin{equation}
c_{f}^2 = \frac{1}{2} \left( c_s^2 + v_A^2 
+ \sqrt{(c_s^2 + v_A^2)^2 - 4 c_s^2 v_{A,n}^2} \right),
\end{equation}

The density distribution reveals subtle variations along the equatorial plane, indicating different stages of departure from perfect equatorial symmetry, which was observed for model \texttt{RUN0}. The influence of magnetic stresses on the flow dynamics is apparent on the shock location and structure of the post-shock region. The shock front exhibits mild distortions and asymmetry, particularly noticeable in the third column of Fig.~\ref{fig:beta1e5}. The corresponding plasma $\beta$ distribution, shown in the bottom row, indicates that although the flow remains gas-pressure-dominated ($\beta > 1$), the structure appears turbulent with some voids of low-density regions in the post-shock disk. This is mainly attributed to the temporal behavior of the shock front. The magnetic field accumulated near the post-shock region tends to push the shock surface outward, while the inflowing matter exerts an opposing ram pressure that drives it inward toward the accreting center. \\

\begin{figure*}[h]
	\begin{center}       \includegraphics[width=0.5\textwidth]{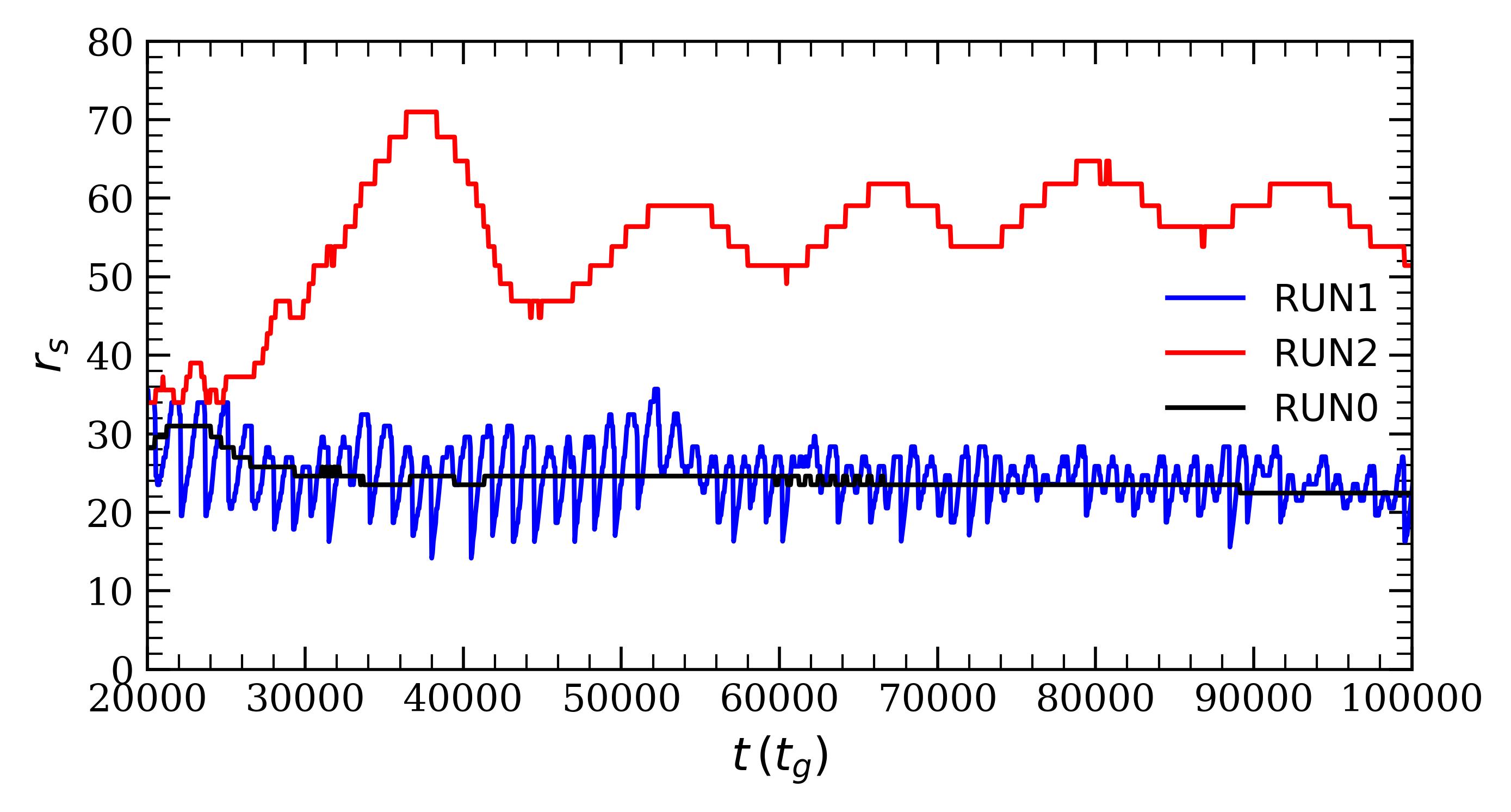}
        \caption{Temporal variation in the shock front position for different simulation runs, as indicated in the labels.}
        \label{fig:shocktime}
        \end{center}
\end{figure*}

To illustrate how these competing effects contribute to the destabilization of the shock front, we examined the temporal evolution of the shock location on the equatorial plane, as shown in Fig.~\ref{fig:shocktime}. The shock exhibits clear oscillatory behavior over time. For comparison, we also show the temporal variation in the shock front for model \texttt{RUN0}, in which the shock remains completely stable, displaying negligible changes in its position throughout the simulation.

 \begin{figure}
	\begin{center}       \includegraphics[width=\textwidth]{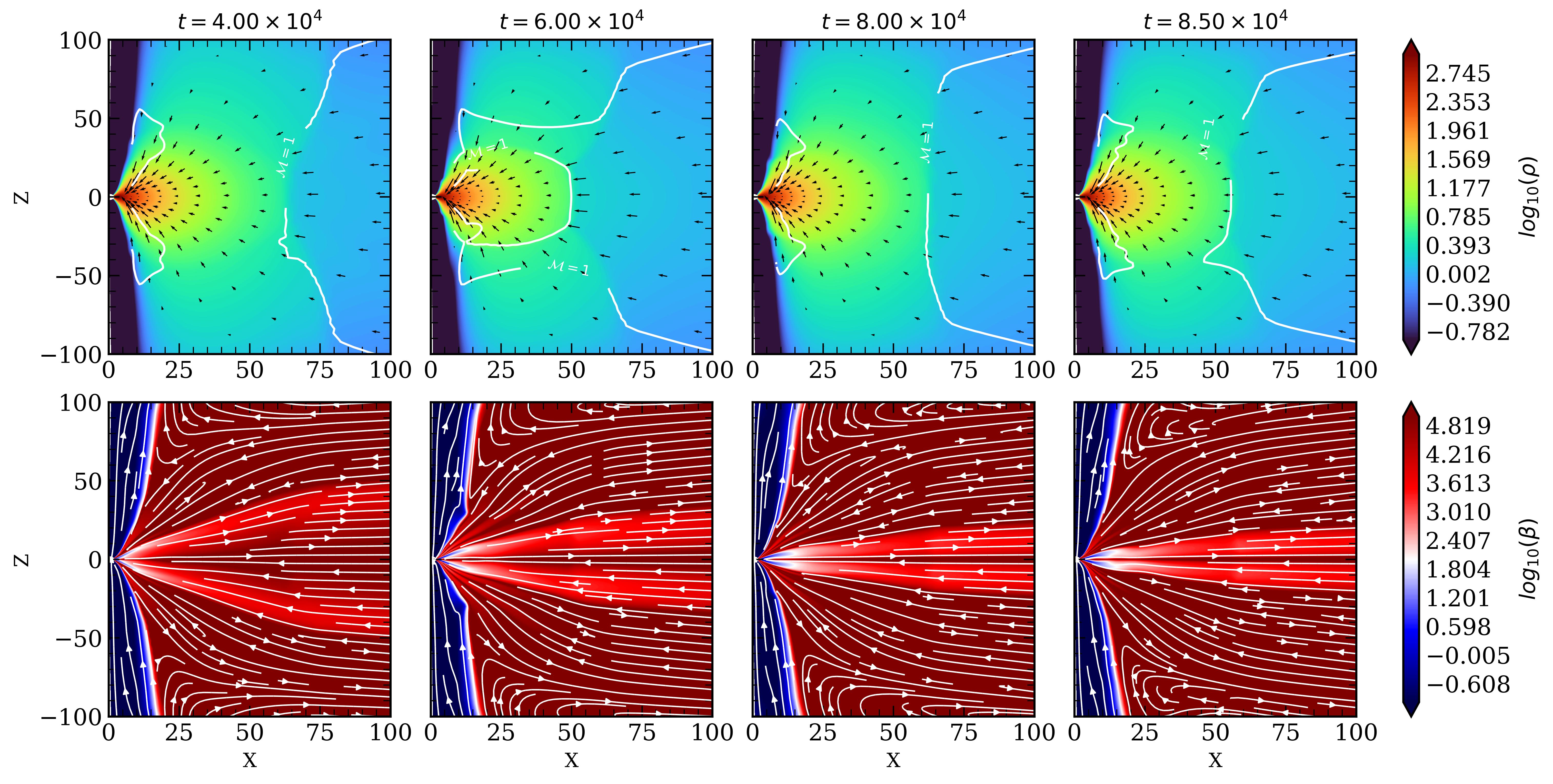}
        \caption{Same as Fig. \ref{fig:beta1e5} but for model \texttt{RUN2}. Magnetic field lines are plotted in the bottom panel.}
        \label{fig:mod4}
        \end{center}
\end{figure}

In model \texttt{RUN2}, the shock front is displaced farther out, as shown in the top panels of Fig. \ref{fig:mod4}. The structure of the post-shock region differs noticeably from that in \texttt{RUN0}, with the mean shock location occurring at significantly larger radii. In the bottom row, we show the magnetic field lines. The accreting material pulls the field lines inward as it falls. However, the rotating flow does not extend into the region close to the axis; the field lines there largely preserve their initial vertical configuration.
Additionally, a comparison with Figs.~\ref{fig:beta1e5} and \ref{fig:hydro} shows that, in this case, the post-shock region becomes more vertically extended, even though the associated density jump is not particularly large. The bottom row shows that there are some regions confined close to the equatorial plane where plasma $\beta$ reduces to the values of $1$--$10$. The time series for the location of the shock front plotted in Fig.~\ref{fig:shocktime} shows that the oscillation amplitude of the shock front increases, while the oscillation frequency decreases. This reduction in frequency is a consequence of the larger size of the post-shock region, which lengthens the dynamical timescale of the shock oscillations. \\

 \begin{figure}
	\begin{center}       \includegraphics[width=\textwidth]{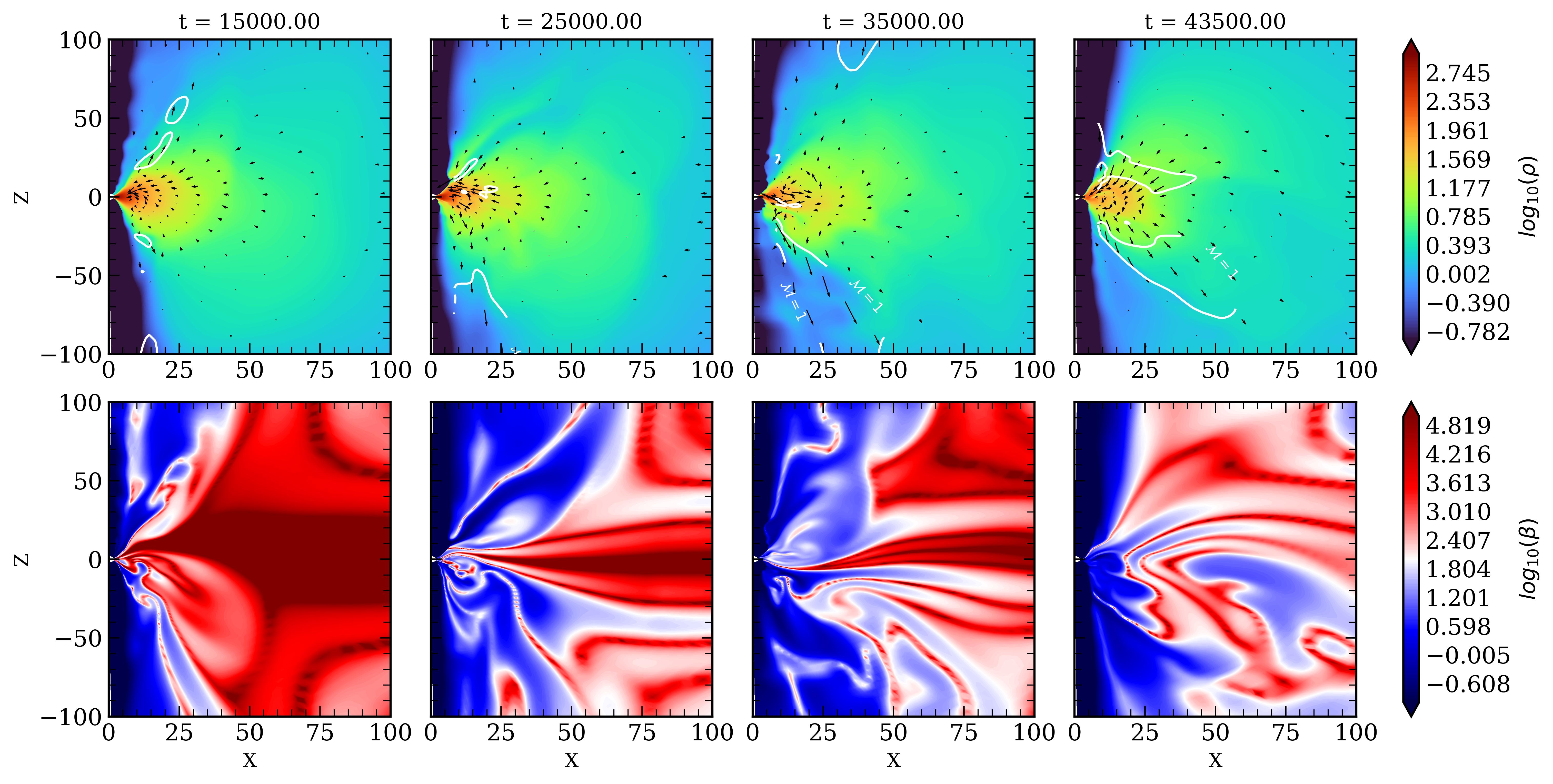}
        \caption{Same as Fig. \ref{fig:beta1e5} but for model \texttt{RUN3}.}
        \label{fig:mod3}
        \end{center}
\end{figure}

Finally, regarding the evolution of model \texttt{RUN3} with $\beta_0 = 10^{3}$, the morphology of the accretion flow undergoes a significant transformation, as shown in Fig. \ref{fig:mod3}; we do not see the formation of a steady shock in this case. The magnetic field becomes dynamically dominant, and in the inner regions we see the formation of turbulent structures with lower plasma $\beta$. 
The equatorial symmetry seen in previous runs is largely disrupted, giving rise to asymmetric structures, as seen in the second and fourth columns of Fig. \ref{fig:mod3}. Moreover, we observe that a fraction of the accreting matter is pushed outward from the vicinity of the black hole, a behavior primarily attributed to the enhanced magnetic pressure in the inner accretion region. \\

\begin{figure*}[h]
\centering
\begin{minipage}{0.98\textwidth}
    \centering
    \includegraphics[width=\textwidth]{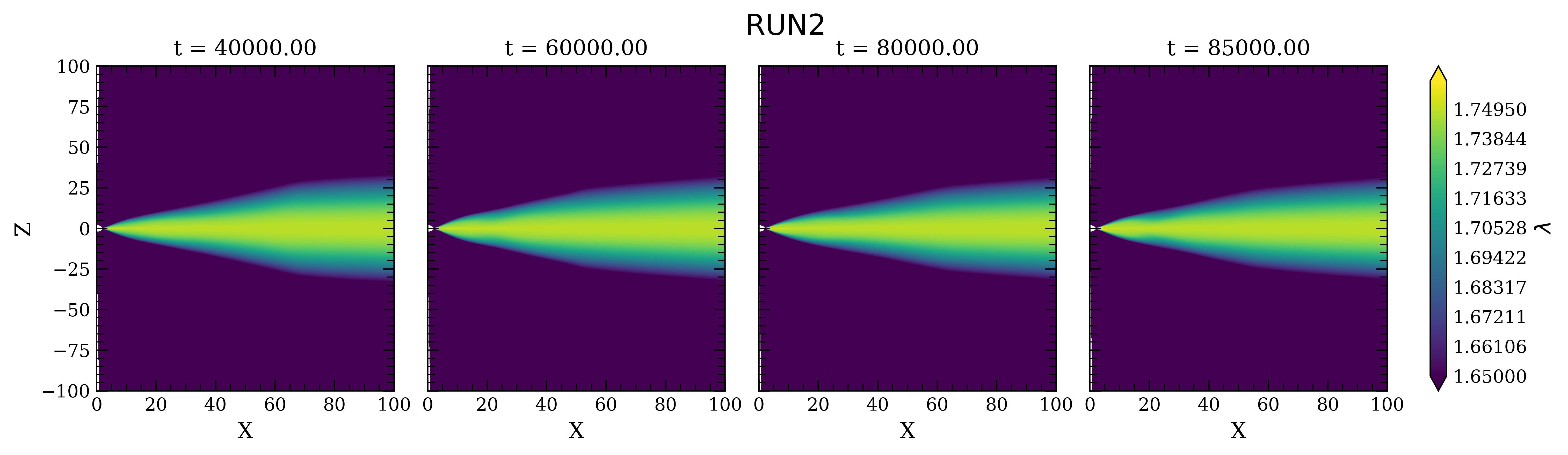}
\end{minipage}
\vspace{1em} 
\begin{minipage}{0.98\textwidth}
    \centering
    \includegraphics[width=\textwidth]{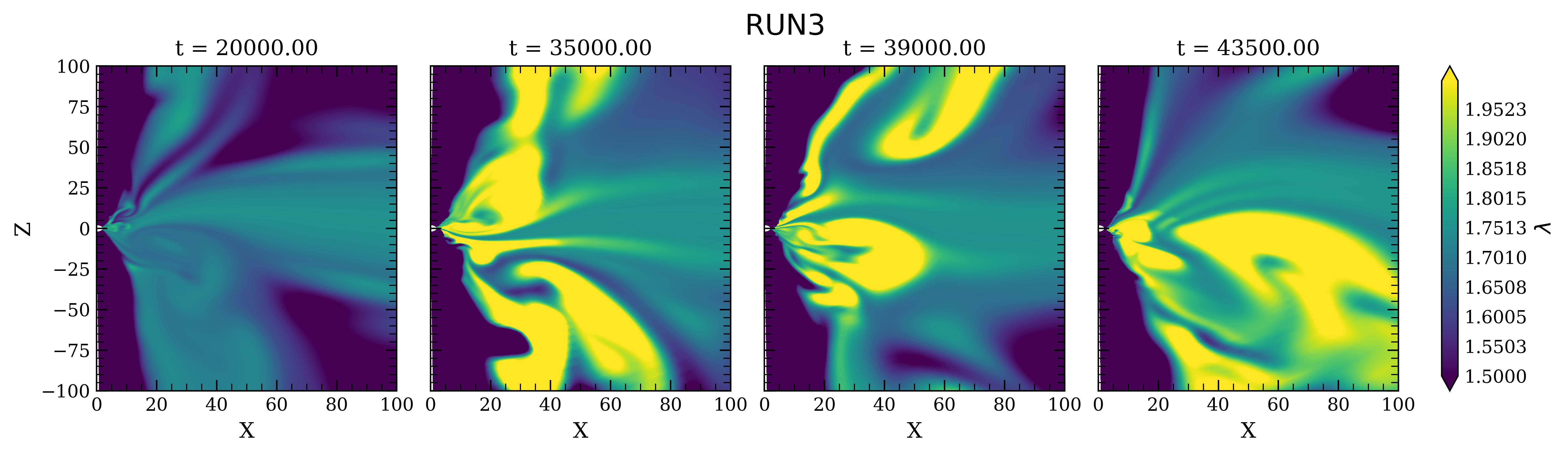} 
\end{minipage}

\caption{Snapshots of the angular momentum distribution for models \texttt{RUN2} and \texttt{RUN3}.}
\label{fig:lamda}
\end{figure*}

In Fig.~\ref{fig:lamda}, the angular momentum distribution at different times is shown for models \texttt{RUN2} and \texttt{RUN3}. In model \texttt{RUN3}, we observe clear signatures of large-scale angular momentum transport away from the equatorial plane. Similar behavior was reported by \citet{2024MNRAS.528.3964D}, where such redistribution was driven by viscous stresses within the disk. In the present case, however, no explicit viscosity is included in the simulations; consequently, angular momentum transport occurs through Maxwell stresses and, where present, Reynolds stresses associated with turbulent motions, rather than through viscous stresses.

In contrast, model \texttt{RUN2} shows no comparable large-scale vertical gradients; the angular momentum remains highest within a well-confined disk region. To check this, we also examined the MRI quality factor $Q$ \citep{2004ApJ...605..321S} for these models and confirmed that it is sufficiently resolved in \texttt{RUN3} ($Q>30$), whereas in \texttt{RUN2} it remains lower ($Q<10$ in the disk regions). The laminar nature of the accretion flow and the low value of $Q$ are similar to those found in the Standard and Normal Evolution (SANE) model of accretion \citep{2012MNRAS.426.3241N,2019ApJS..243...26P}.
The absence of extended angular momentum transport in \texttt{RUN2} suggests that the magnetic field is not strong enough to drive significant angular momentum transport, but the magnetic pressure and tension forces are sufficient to introduce an imbalance in the net forces, which drives the observed shock oscillations.

\subsection{Effect of Shock Oscillation}
\label{sec:lum}
The post-shock region is hotter and denser in comparison to the pre-shock region; hence it emits high-energy photons \citep{2020A&A...642A.209S}. The oscillations in shock location are expected to be reflected in the emitted radiation. We investigate this effect by analyzing the time series of luminosity calculated for different sets of simulations. We estimate the bremsstrahlung and synchrotron losses through a posteriori calculations as the representation of radiative processes in the accretion disk. The emissivity resulting from bremsstrahlung (measured in ${\mathrm{erg}~\mathrm{cm}^{-3}~\mathrm{s}^{-1}}$) is described by \cite{1973blho.conf..343N} and
given as,
\begin{equation}
Q_{\rm br}=1.4\times10^{-27} n^2_e\sqrt{T_e}\left(1+4.4\times10^{-10}T_e\right)
 \label{eq:bremm}
\end{equation}
and the emissivity due to the synchrotron (in ${\mathrm{erg}~ {\rm cm}^{-3}~ {\rm s}^{-1}}$) is given by \cite{1983JBAA...93R.276S},
\begin{equation}
Q_{\rm syn}=\frac{16}{3} \frac{e^2}{c} \left( \frac{eB}{m_e c} \right)^2 \left(\frac{k_bT_{e}}{m_ec^2}\right)^2 n_e 
 \label{eq:synchro}
\end{equation}
At any time $t$, the total luminosity is the volume integral of the emissivities given as
\begin{equation}
 L=\int (Q_{\rm syn}+Q_{\rm br}) dV  
\label{eq:cool}
\end{equation}
We obtain the luminosity profiles for a black hole mass, $M=10M_\odot$. To calculate the density, we assume an accretion rate $\dot{M}=0.1\,M_{Edd}$, where $M_{Edd}=1.44\times10^{17}(M/M_\odot)g/s$ is the Eddington accretion rate. The density at the outer boundary (taken as $\rho=1$ in code units) can be computed in physical units using the relation $\rho=\dot{M}/(4\pi\,v_{inj}r_{inj}^2 {\rm cos}\Delta)$, with $\Delta$ being the half-angular extent of the disk. We use the single-temperature assumption to calculate electron and ion temperatures $T_e=T_p=\frac{pm_p}{\rho k_b}$.

\begin{figure*}[h]
\centering
\begin{minipage}[t]{0.48\textwidth}
    \centering
    \includegraphics[width=\textwidth]{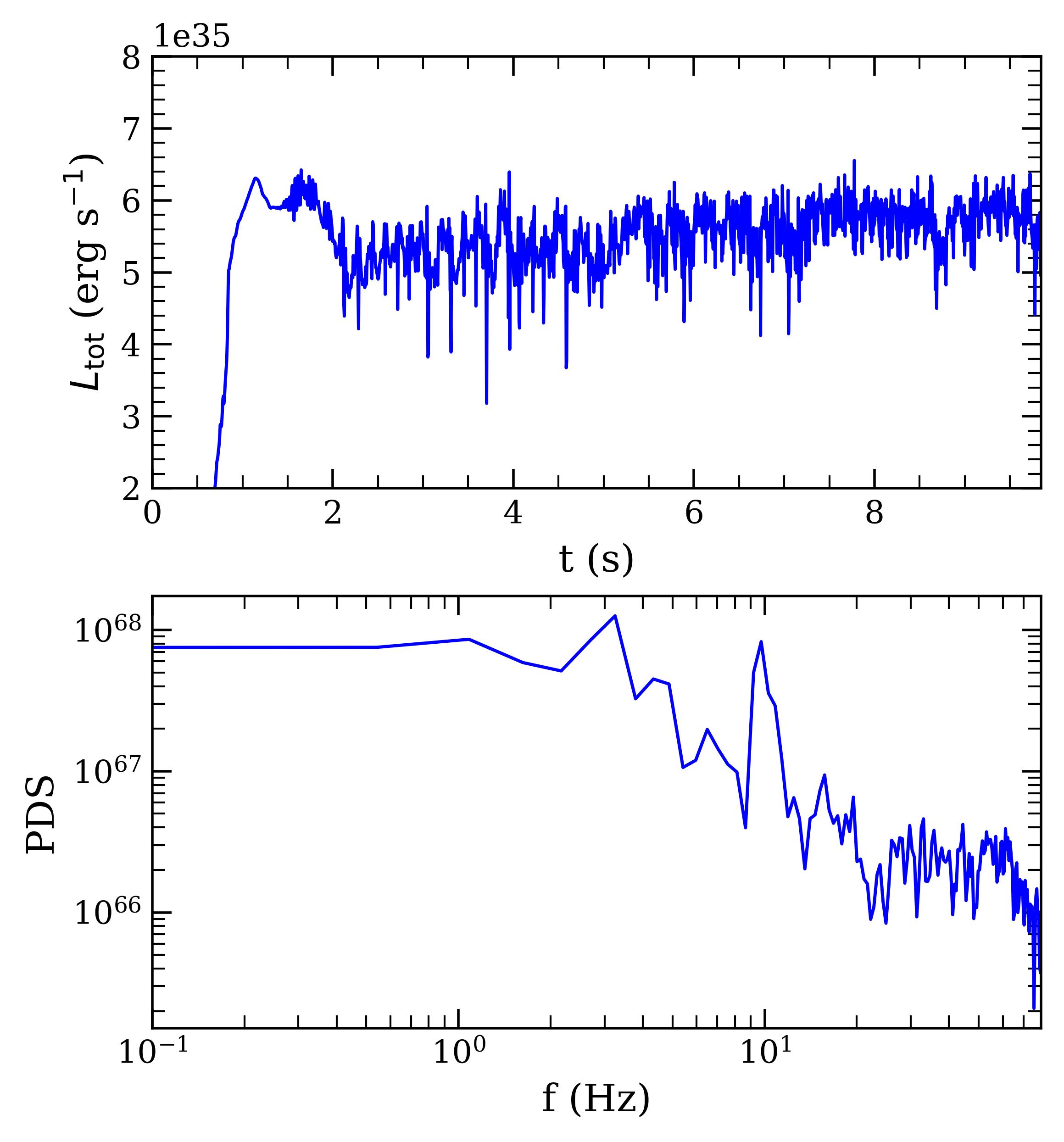}
\end{minipage}
\hfill
\begin{minipage}[t]{0.48\textwidth}
    \centering
    \includegraphics[width=\textwidth]{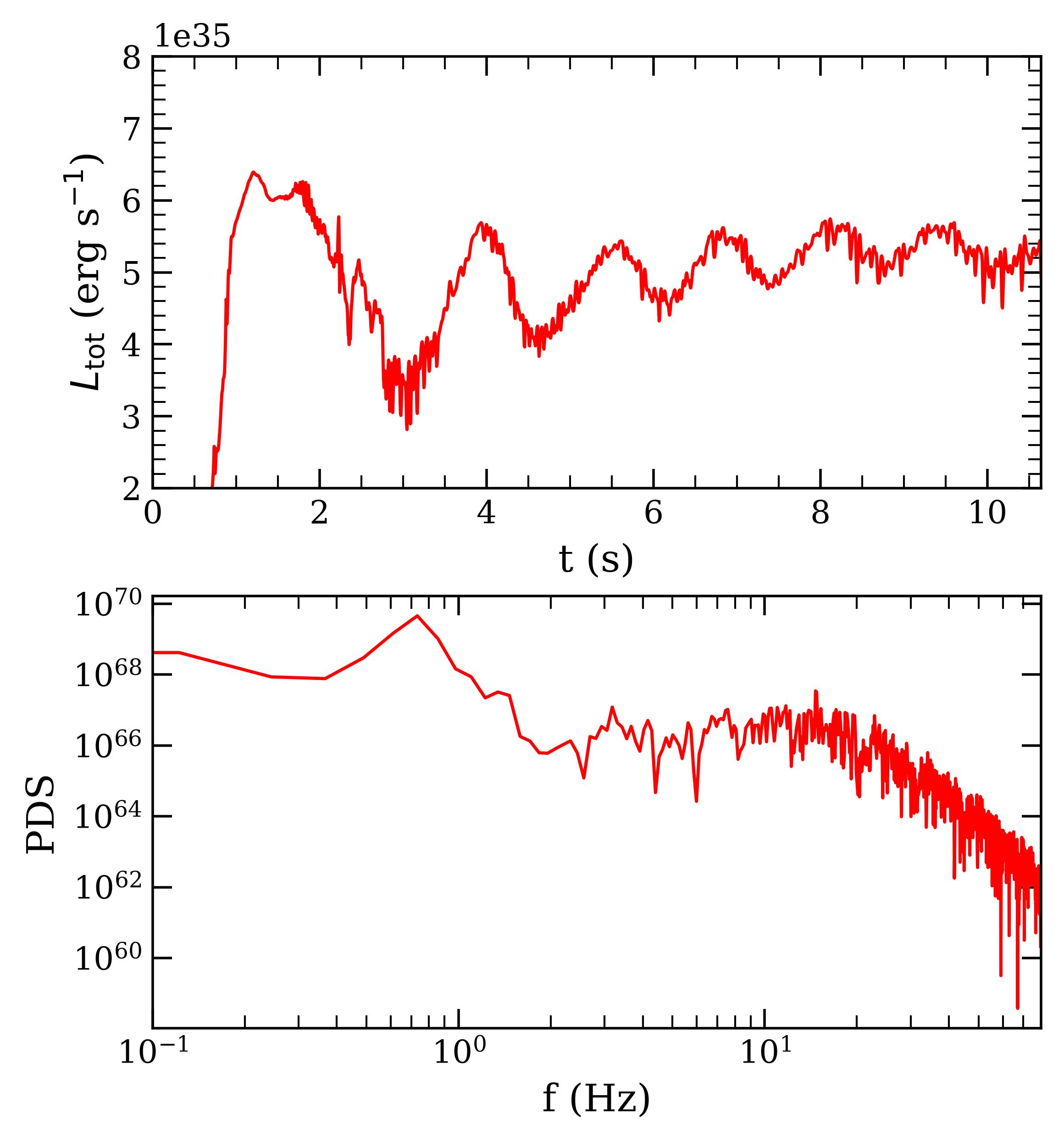} 
\end{minipage}

\caption{Time series of luminosity in units of ($\rm erg\,s^{-1}$) and corresponding power spectra for model \texttt{RUN1} (left column) and \texttt{RUN2} (right column).}
\label{fig:lumpsd}
\end{figure*}

The oscillatory behavior of the shock front is directly imprinted on the luminosity of the accretion flow. In the top panels of Fig.~\ref{fig:lumpsd}, we present the time evolution of the bolometric luminosity for models \texttt{RUN1} (left) and \texttt{RUN2} (right), with the corresponding power density spectra (PDS) shown in the lower panels. Model \texttt{RUN1} displays much more rapid luminosity fluctuations than \texttt{RUN2}, consistent with the faster shock oscillations in that model. A shock forming closer to the black hole heats the post-shock region more efficiently than one located farther out; thus, outward excursions of the shock naturally produce lower-luminosity phases, giving rise to the observed maxima and minima.

In model \texttt{RUN2}, the luminosity varies more slowly, with well-separated peaks and troughs, though small-scale fluctuations remain superimposed on the broader modulation. To identify QPOs associated with these variations, we computed the PDS of the luminosity signal, shown in the bottom panels. The PDS of \texttt{RUN1} exhibits two prominent peaks near 10~Hz and 2~Hz. Additional features appear at 15~Hz and 18~Hz, but with power at least an order of magnitude lower than the low-frequency peaks. For model \texttt{RUN2}, the slower luminosity modulation produces a QPO at 0.75~Hz.

These results demonstrate that oscillating shock fronts can naturally generate QPOs over a range from sub-Hz to a few tens of Hz, depending on the mean location of the shock. Oscillations occurring closer to the black hole are faster and lead to rapid luminosity variations, whereas outwardly displaced shocks have longer dynamical timescales, reducing the characteristic oscillation frequency.

\begin{figure}[ht]
\centering
\begin{minipage}{0.85\textwidth}
    \centering
    \includegraphics[width=\textwidth]{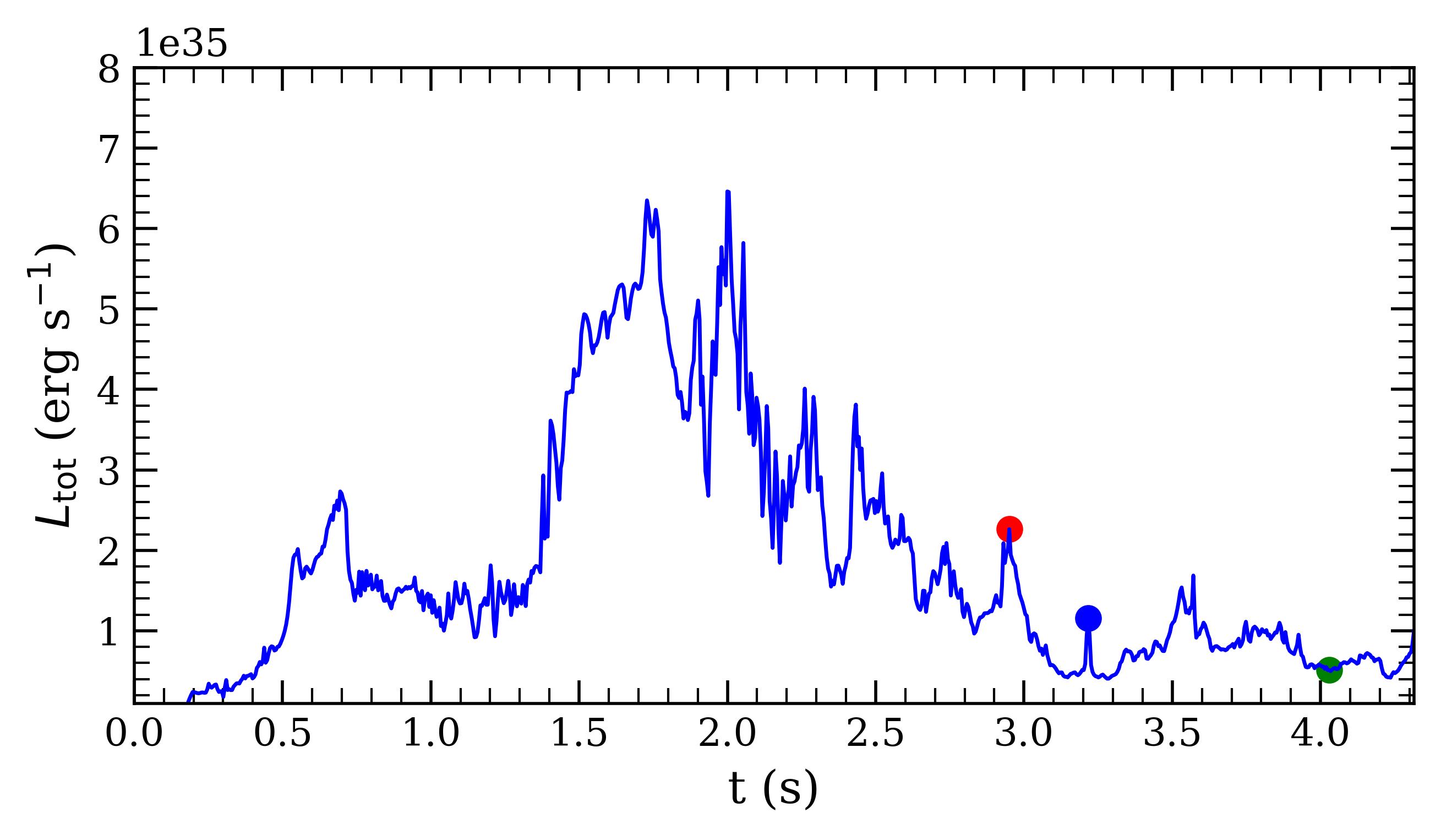}
\end{minipage}
\vspace{1em} 
\begin{minipage}{0.85\textwidth}
    \centering
    \includegraphics[width=\textwidth]{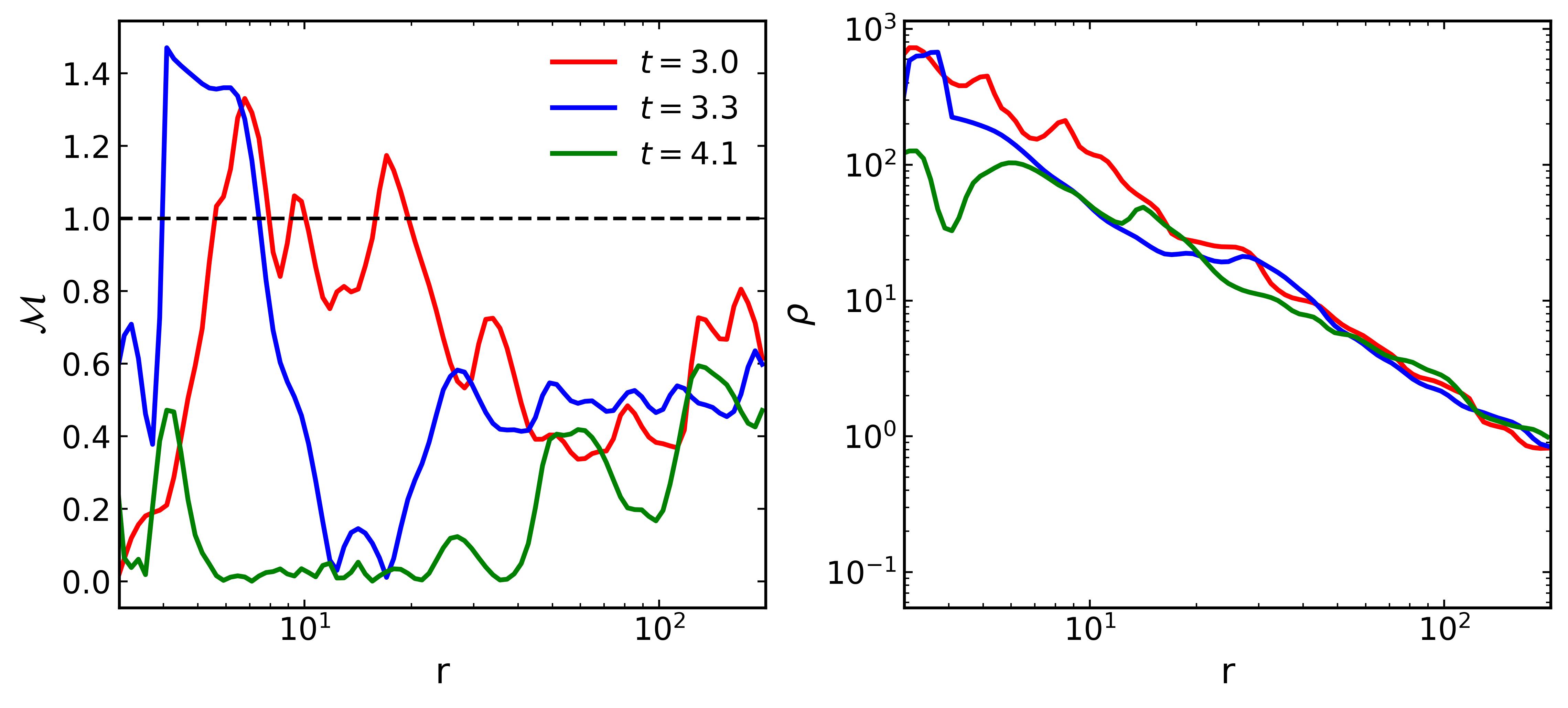}
\end{minipage}

\caption{Variation in luminosity for \texttt{RUN3} in top panel and bottom panels show equatorial plane profiles of Mach number and density plotted at time stamps mentioned in legend.}
\label{fig:beta3diag}
\end{figure}
The time series of luminosity for model \texttt{RUN3} is shown in Fig.~\ref{fig:beta3diag}. In this case, no steady shock forms throughout the simulation, and as a result, the bolometric luminosity remains noticeably lower than in models \texttt{RUN2} and \texttt{RUN1}. This reduced emission reflects the absence of a persistent post-shock region capable of sustained heating. Although no stable shock is present, we do observe transient shock structures that momentarily appear in the flow and produce brief luminosity spikes. The bottom panel of Fig.~\ref{fig:beta3diag} displays the radial profiles of Mach number and density, confirming the formation of these intermittent shocks; the corresponding spikes in luminosity are marked using matching colors. For comparison, profiles from epochs without any shock are also plotted, corresponding to the low-luminosity states.

Because these shock episodes are sporadic and short-lived, they fail to produce any coherent or quasi-periodic modulation. Consequently, the luminosity curve shows no well-defined oscillatory behavior. This outcome is fully consistent with expectations: in the absence of a stable shock oscillation region, the characteristic timescales required for QPO formation do not arise, leading instead to a predominantly low-luminosity, weakly variable state punctuated only by occasional, short-duration enhancements by a factor of 2--3.

\subsection{Three-Dimensional Simulation}

\begin{figure}
	\begin{center}       \includegraphics[width=\textwidth]{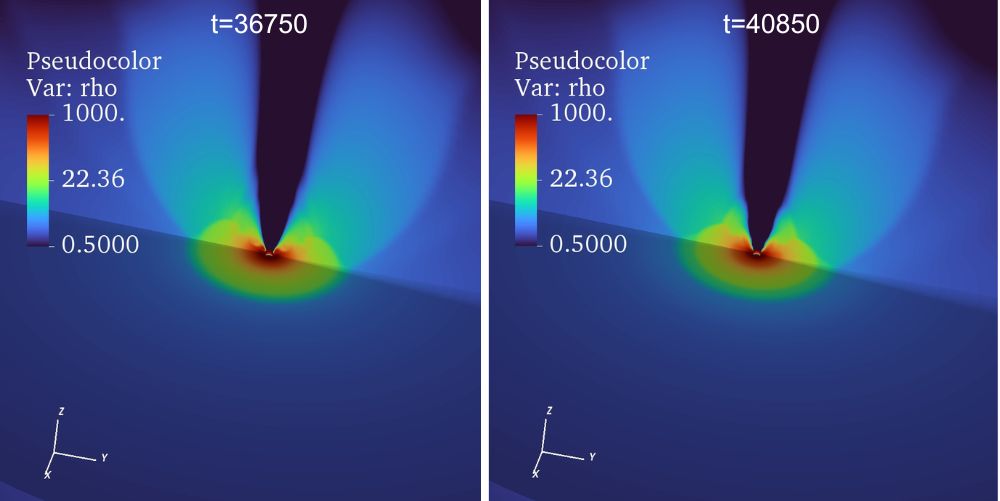}
        \caption{Density distribution for 3D simulation of \texttt{RUN1}.}
        \label{fig:3drho}
        \end{center}
\end{figure}

Recently, \citet{2023MNRAS.519.4550G} demonstrated that in three-dimensional simulations the post-shock region can develop a standing accretion shock instability (SASI), triggered under the influence of non-axisymmetric perturbations. Motivated by this result, we performed fully three-dimensional simulations for model \texttt{RUN1}. We do not impose any external perturbations as in our magnetized setup, the shock already undergoes intrinsic oscillations driven by magnetic pressure, which can naturally seed non-axisymmetric structure. Such features would be suppressed in two-dimensional simulations because of the enforced axisymmetric boundary conditions. The three-dimensional density structure, displayed at two separate time stamps, is shown in Fig.~\ref{fig:3drho}. The structure is visualized using slices taken at the planes $z = 0$ and $y = 0$. Due to the non-zero angular momentum of the flow, the region surrounding the polar axis contains very little matter. The density profiles shown at two different times indicate variations in the post-shock region, driven by oscillations in the shock front; however, the disk retains an overall symmetric structure.

\begin{figure*}[h]
\centering
\begin{minipage}[t]{0.48\textwidth}
    \centering
    \includegraphics[width=\textwidth]{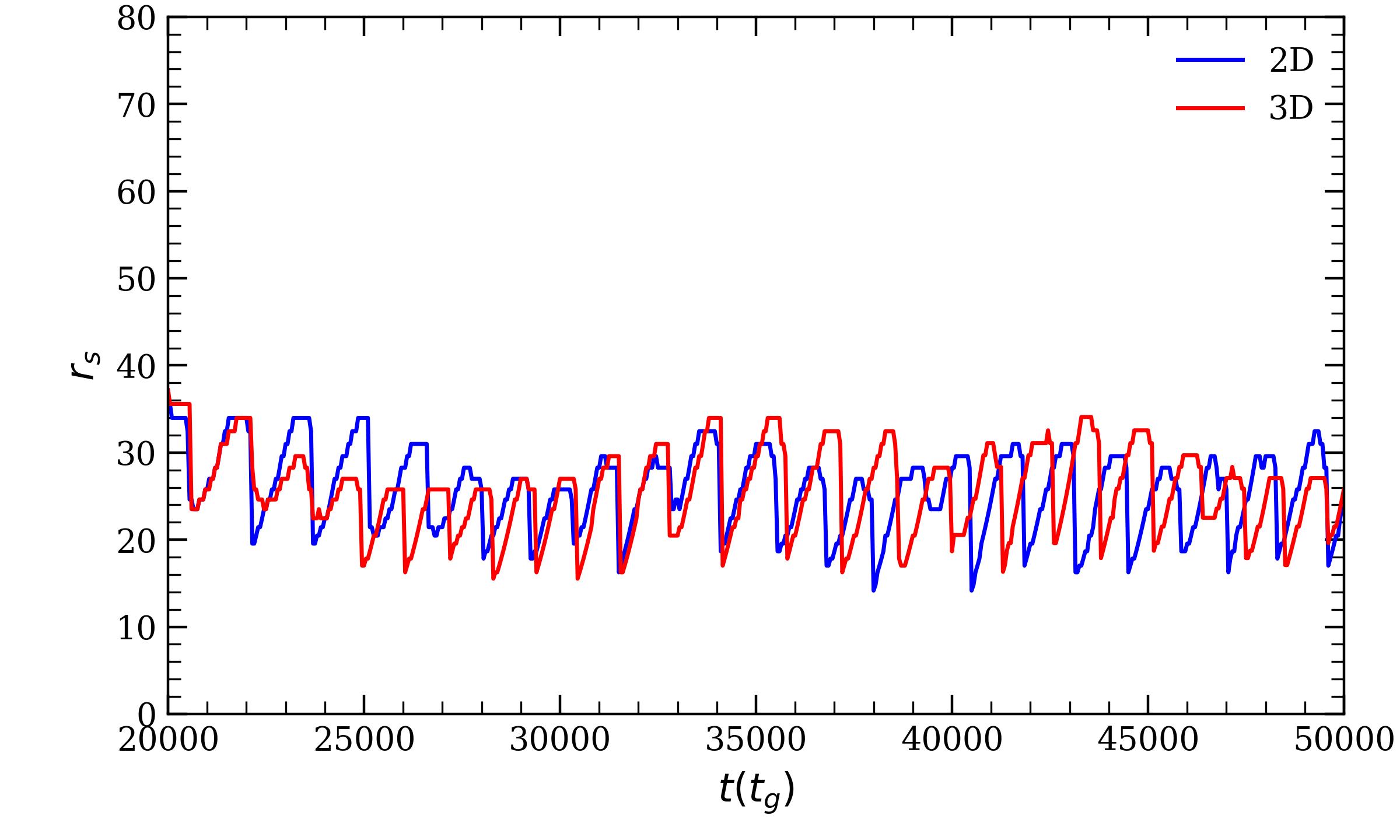}
\end{minipage}
\begin{minipage}[t]{0.48\textwidth}
    \centering
    \includegraphics[width=\textwidth]{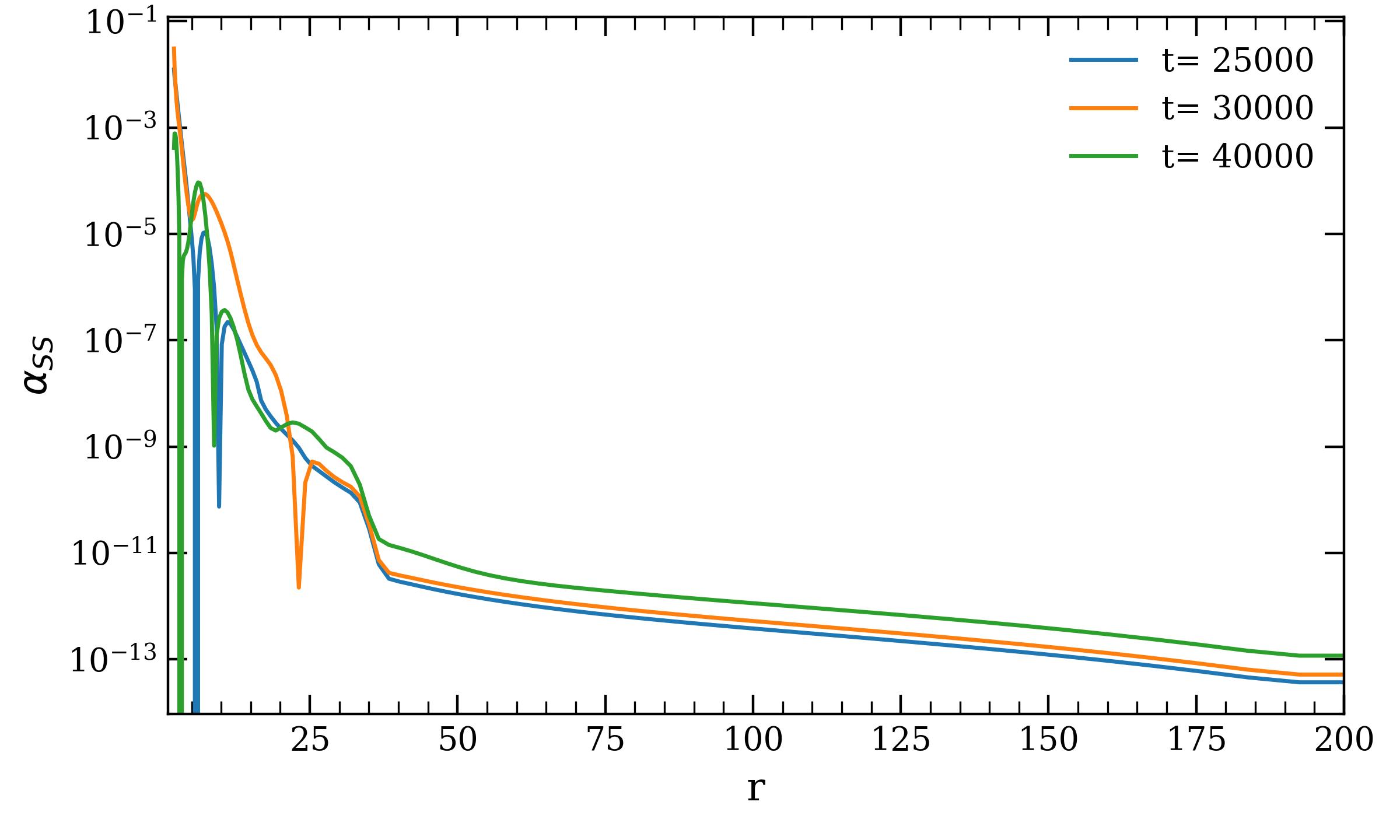}
\end{minipage}
\caption{(Left panel) Comparison between shock positions in the 2D and 3D runs. The (right panel) shows the radial variation in the viscosity parameter.}
\label{fig:2d3d}
\end{figure*}

The comparison of shock oscillation with the 2D run shown in Fig. \ref{fig:2d3d} highlights that the oscillation frequency is almost similar but the shock amplitude differs in comparison to the 2D run.

We also calculate the effective viscosity parameter from Maxwell and Reynolds stresses. 

The Reynolds stress is computed from velocity fluctuations:
\begin{equation}
    T^{\rm Reynolds}_{r\phi} = \rho\, v_r' v_\phi',
\end{equation}
where the fluctuating velocities are defined as
\begin{equation}
    v_r' = v_r - \langle v_r \rangle_\phi, \qquad
    v_\phi' = v_\phi - \langle v_\phi \rangle_\phi.
\end{equation}
Here, $\langle \cdot \rangle_\phi$ denotes an azimuthal average at fixed $(r,\theta)$.  

The Maxwell stress is computed directly from the components of the magnetic field:
\begin{equation}
    T^{\rm Maxwell}_{r\phi} = -\,\frac{B_r B_\phi}{4\pi}.
\end{equation}

The total $r\phi$ component of the stress is
\begin{equation}
    T_{r\phi} = T^{\rm Maxwell}_{r\phi} + T^{\rm Reynolds}_{r\phi},
\end{equation} and the Shakura--Sunyaev viscosity parameter is
\begin{equation}
    \alpha_{ss} = \frac{T_{r\phi}}{P}.
\end{equation}

To characterize radial profiles, we extract $\alpha$, $T^{\rm Maxwell}_{r\phi}$, 
and $T^{\rm Reynolds}_{r\phi}$ along the equatorial plane ($\theta = \pi/2$).  
We further average over the azimuthal angle $\phi$:
\begin{equation}
    {\alpha}_{ss}(r) = \frac{1}{2\pi}\int_0^{2\pi} \alpha(r, \theta=\pi/2, \phi)\,d\phi,
\end{equation}
The radial variation of $\alpha_{ss}$ is plotted in the right panel of Fig. \ref{fig:2d3d}. The maximum value of $\alpha_{ss}$ is of the order of $10^{-2}$. Some negative spikes appear in the $\alpha$ distribution as a result of local turbulent fluctuations. Since model \texttt{RUN1} has a low magnetic field, the expected $\alpha$ values are relatively small. As the accretion is advection-dominated, any small-scale fluctuations will be advected towards the inner region before they can grow and lead to a turbulent state that enhances angular momentum transport; hence the outer regions of the disk ($r>10r_g$) show a very low value of $\alpha$.
In contrast, models with higher magnetic fields, such as \texttt{RUN3}, exhibit more turbulent behavior in 2D simulations, as the enhanced
magnetic pressure also works against gravity and slows down the infalling matter. This leads to a longer infall timescale, allowing for growth of turbulence, and the dynamical behavior resembles that of cases with higher-viscosity models, such as those of \citet{2025ApJ...994...48D}.

\begin{figure*}[h]
\centering
\begin{minipage}[t]{0.48\textwidth}
    \centering
    \includegraphics[width=\textwidth]{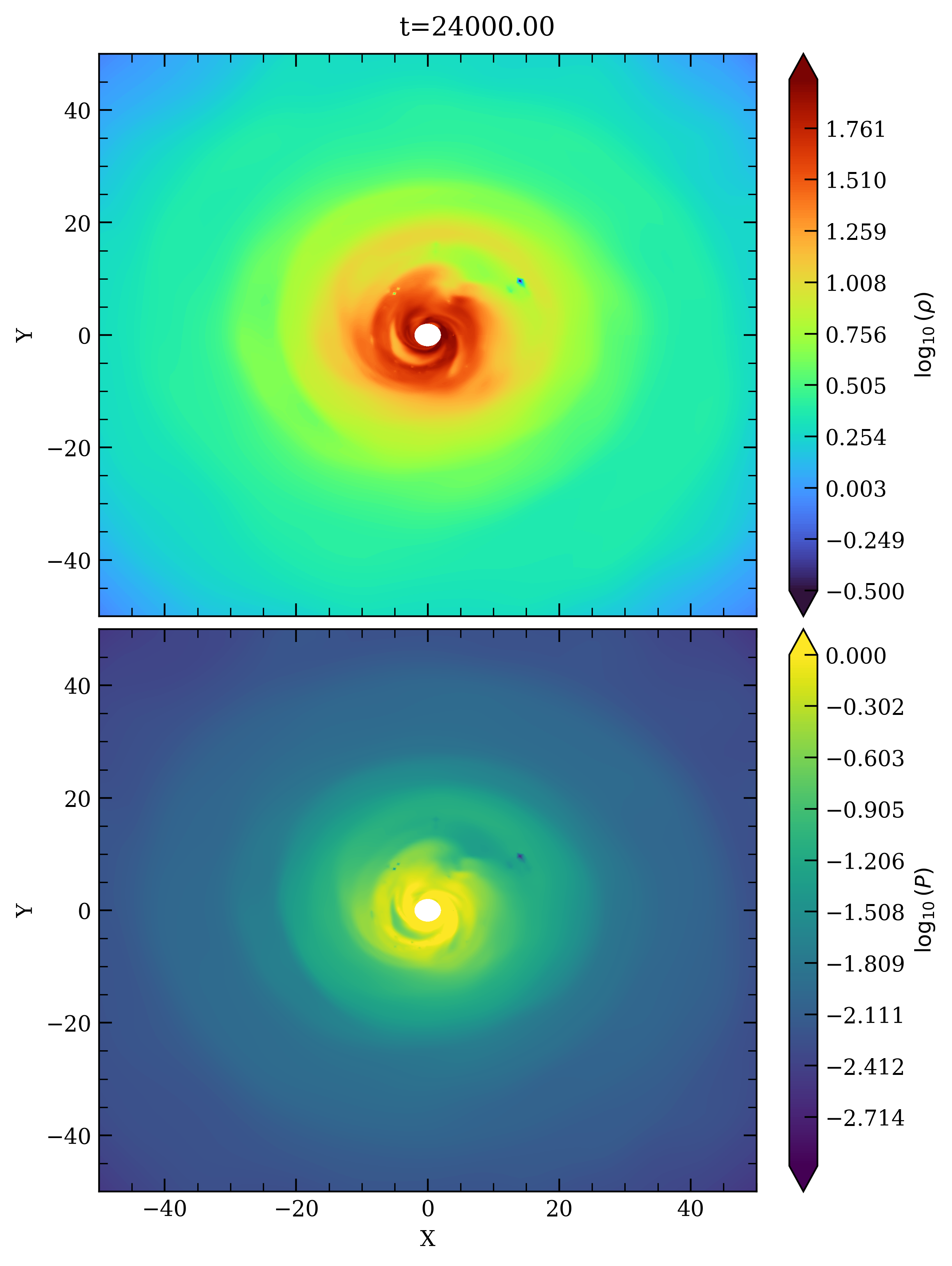}
\end{minipage}
\begin{minipage}[t]{0.48\textwidth}
    \centering
    \includegraphics[width=\textwidth]{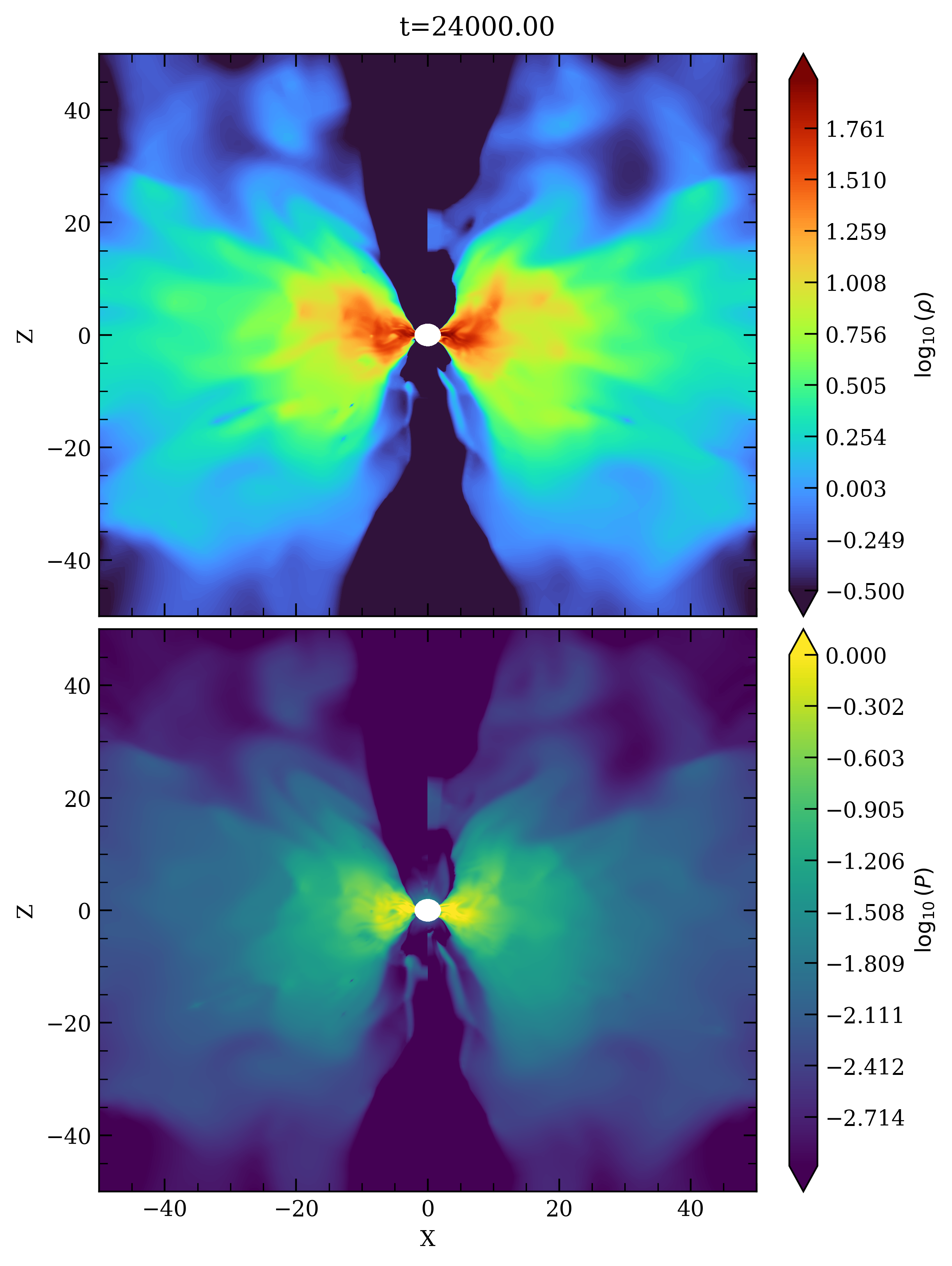}
\end{minipage}
\caption{Density and pressure distributions from the 3D simulation of \texttt{RUN3}, shown in the equatorial plane (left column) and meridional plane (right column).}
\label{fig:3drun3}
\end{figure*}

\begin{figure*}[h]
\centering
\begin{minipage}[t]{0.48\textwidth}
    \centering
    \includegraphics[width=\textwidth]{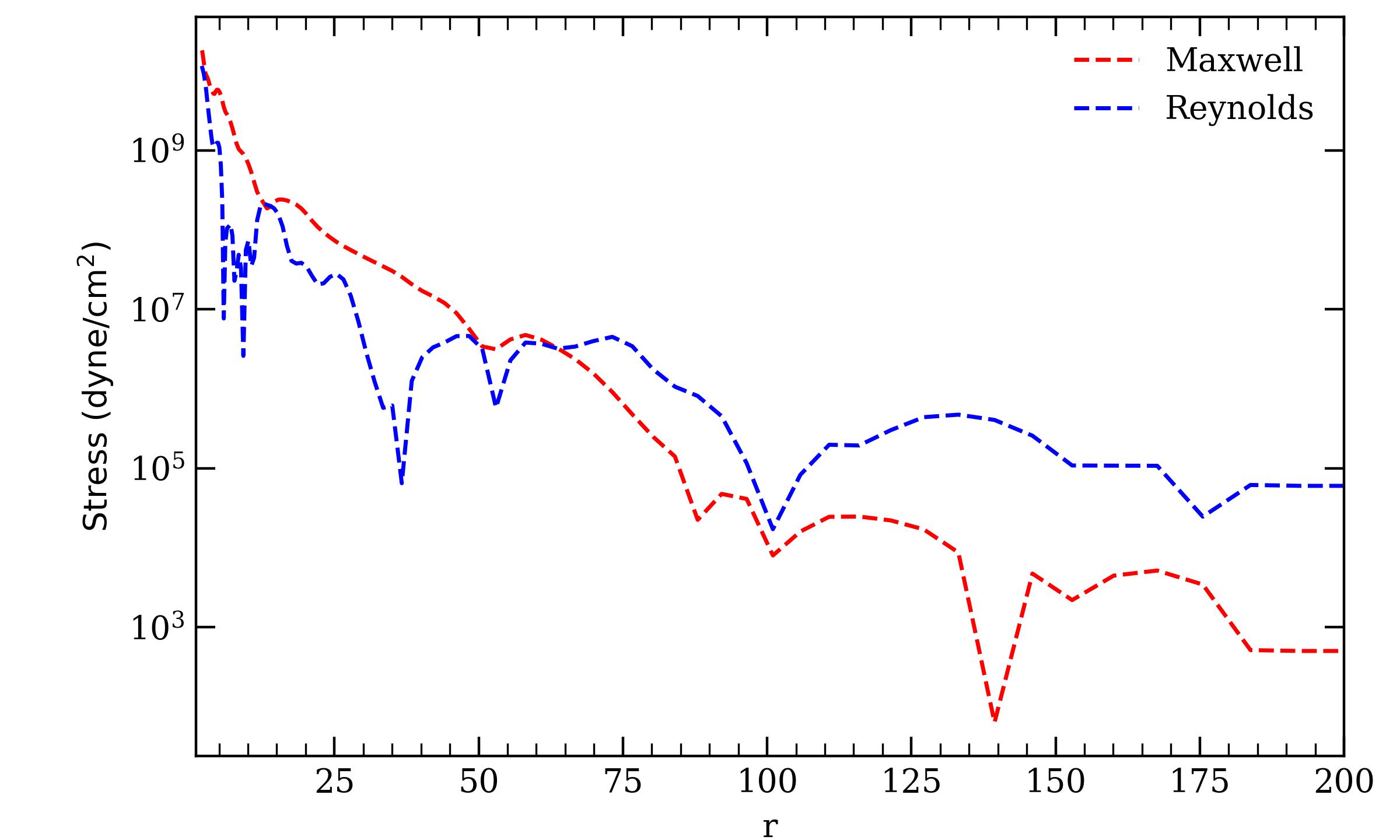}
\end{minipage}
\begin{minipage}[t]{0.48\textwidth}
    \centering
    \includegraphics[width=\textwidth]{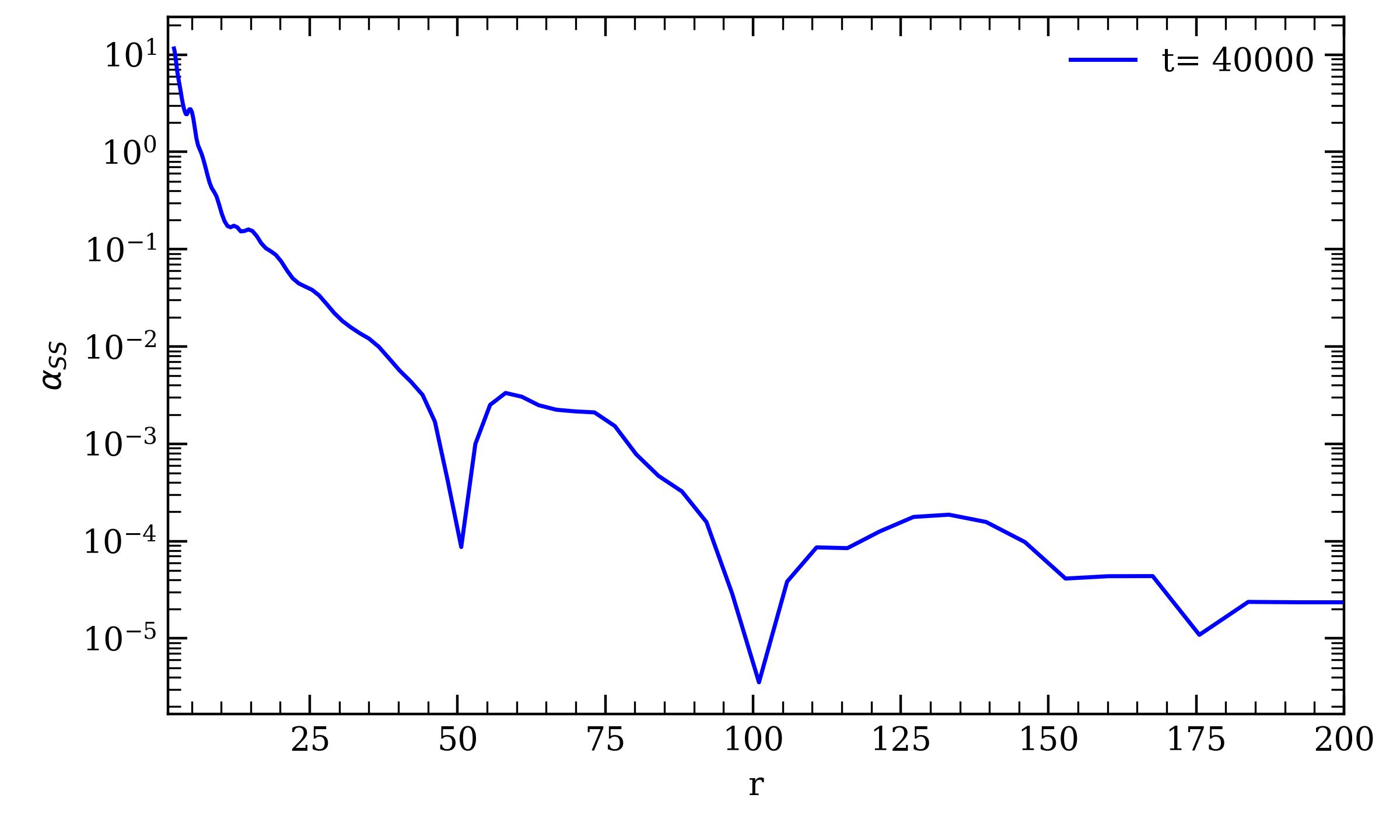}
\end{minipage}
\caption{(Left panel) Variation in Reynolds and Maxwell stress for \texttt{RUN3}; the (right panel) shows the radial variation in the Shakura--Sunyaev viscosity parameter.}
\label{fig:diagnos_r3}
\end{figure*}

We also performed a 3D simulation for \texttt{RUN3} in order to explore whether the flow shows turbulent behavior in 3D. The density and pressure distributions for the simulations are plotted in Fig. \ref{fig:3drun3}.
The equatorial snapshot reveals a rotationally supported disk characterized by a dense, high-pressure inner region and tightly wound spiral features, which fragment into irregular, clumpy, over-dense, and filamentary features, indicating turbulent mixing rather than a steady, symmetric spiral pattern. Small-scale density inhomogeneities and pressure fluctuations are mostly confined within the inner regions of the disk. The meridional slice likewise departs from bilateral symmetry; the polar cavities show uneven opening angles, patchy density structures, and localized knots, implying intermittent, anisotropic mass loading and turbulent entrainment along the funnel walls. These features were not present in the weakly magnetized case \texttt{RUN1}, which evolved as a steady, symmetric configuration.\\
To obtain a quantitative measure of turbulence driven by magnetically mediated processes, we present radial profiles of the Reynolds stress, Maxwell stress, and effective viscosity parameter in Fig.~\ref{fig:diagnos_r3}. The first panel shows that the Maxwell stress consistently exceeds, or closely follows, the Reynolds stress at small radii, indicating that correlated magnetic field fluctuations dominate the outward transport of angular momentum, as expected for magnetorotationally driven turbulence. Both stress components decrease rapidly with radius, marking the transition from a highly turbulent inner disk to a more diffuse, nearly laminar outer flow. The second panel shows that $\alpha_{\rm SS}$ attains values of order unity in the innermost region, significantly larger than those in Fig.~\ref{fig:2d3d}, where the flow exhibits no clear signatures of turbulence.

\section{Discussion and Conclusions}
\label{sec:concl}
In this study, we explored the behavior of transonic accretion solutions in the magnetized regime. We performed ideal magnetohydrodynamic (MHD) simulations of accretion flows around a non-rotating black hole using a pseudo-Newtonian potential. The injection parameters for our simulations were derived from steady-state analytical transonic solutions, ensuring consistency between the numerical and analytical frameworks.

Our results demonstrate that accretion flows around black holes can form standing shock fronts as a consequence of the centrifugal barrier experienced by the inflowing matter. We then investigated the evolution of these shocked solutions under the influence of magnetic fields. By fixing the hydrodynamic parameters and systematically varying the magnetic field strength, we were able to study the effect of magnetic forces in a controlled manner. The introduction of magnetic fields produces significant modifications in the overall morphology and dynamics of the flow. Magnetic pressure and tension forces act to distort the shock surface, leading to a more irregular and time-dependent structure compared to the purely hydrodynamic case. Starting with very weak magnetic fields we showed that the global structure of the disk is preserved but the shock front starts to oscillate. As we increase the strength of the magnetic field, the magnetic pressure becomes comparable to the gas pressure, and the shock front exhibits fragmented behavior. The corresponding density and pressure distributions become increasingly asymmetric, indicating the development of magnetically driven instabilities.

Our results suggest that magnetic fields can destabilize an otherwise steady hydrodynamic shock configuration. Since the shock front compresses the inflowing matter and generates a hot, dense post-shock region, it naturally creates a boundary between the cold pre-shock and hot post-shock flows. Consequently, any temporal variations in the position or stability of the shock are expected to manifest as variability in the observed luminosity. This effect was reflected in the simulated bolometric luminosity, where temporal fluctuations trace either the periodic or the transient nature of the shocks, depending on the strength of the magnetic fields. \\

Although in this study we have used a pseudo-Newtonian potential to approximate the gravity of a non-rotating black hole, we expect the qualitative behavior of the flow to remain similar in the full general relativistic regime for a Schwarzschild black hole. The Paczy\'nski--Wiita potential effectively captures a significant portion of the parameter space of transonic and shocked solutions when compared with fully relativistic treatments \cite{2014MNRAS.443.3444K}. 

In the present study, we investigated the stability of shocked solutions by imposing a magnetic field on a flow obtained from a non-magnetized solution. However, analytical studies by \citet{mitra24} have demonstrated that steady-state solutions can also be derived self-consistently within the GRMHD framework. For the future work, we plan to extend this study by including black hole rotation and full general relativistic magnetohydrodynamic (GRMHD) effects to account for frame-dragging and enhanced magnetic coupling near the event horizon. 

Furthermore, incorporating radiative cooling processes and more realistic equations of state will allow a more direct connection between these simulations and observable properties, such as spectral variability and quasi-periodic oscillations (QPOs). Although these effects are significant, they are beyond the scope of this study and will be explored in future work, with results to be reported separately.


\authorcontributions{Conceptualization, R.K.J. and I.C.; Methodology, R.K.J.; Validation, S.D., I.C. and R.A.; Formal analysis, R.K.J. and S.D.; Investigation, R.K.J.; Resources, A.T.; Writing---original draft, R.K.J. and A.T.; Writing---review \& editing, A.T., S.D., I.C. and R.A.; Visualization, R.K.J. and S.D.; Supervision, I.C. All authors have read and agreed to the published version of the manuscript.}

\funding{
This work was supported in part by National Science Foundation (NSF) grants No. PHY-2308242 and No. OAC-2310548 to the University of Illinois
at Urbana-Champaign. A.T. acknowledges support from the National Center for
Supercomputing Applications (NCSA) at the University of Illinois at
Urbana-Champaign through the NCSA Fellows program. This work used Anvil at Purdue University through allocation MCA99S008 from the Advanced Cyberinfrastructure Coordination Ecosystem:
Services \& Support (ACCESS) program, which is supported by National Science
Foundation grants \#2138259, \#2138286, \#2138307, \#2137603, and \#2138296.
This research also used Frontera at TACC through allocation AST20025. Frontera
is made possible by NSF award OAC-1818253. The initial computations and analysis for this article were conducted using the computer cluster at the Nicolaus Copernicus Astronomical Center of the Polish Academy of Sciences (CAMK PAN).
}

\dataavailability{The data underlying this article will be shared on reasonable request
to the corresponding author.}

\acknowledgments{R.K.J. would like to thank Miljenko \v{C}emelji\'{c} for helpful comments during the preparation of this manuscript.}

\conflictsofinterest{The authors declare no conflicts of interest.} 

%


\appendixtitles{} 
\appendixstart
\appendix
\section[One-Dimensional Simulation]{One-Dimensional Simulation}
\label{sec:app1}
\begin{figure}
	\begin{center}       \includegraphics[width=\textwidth]{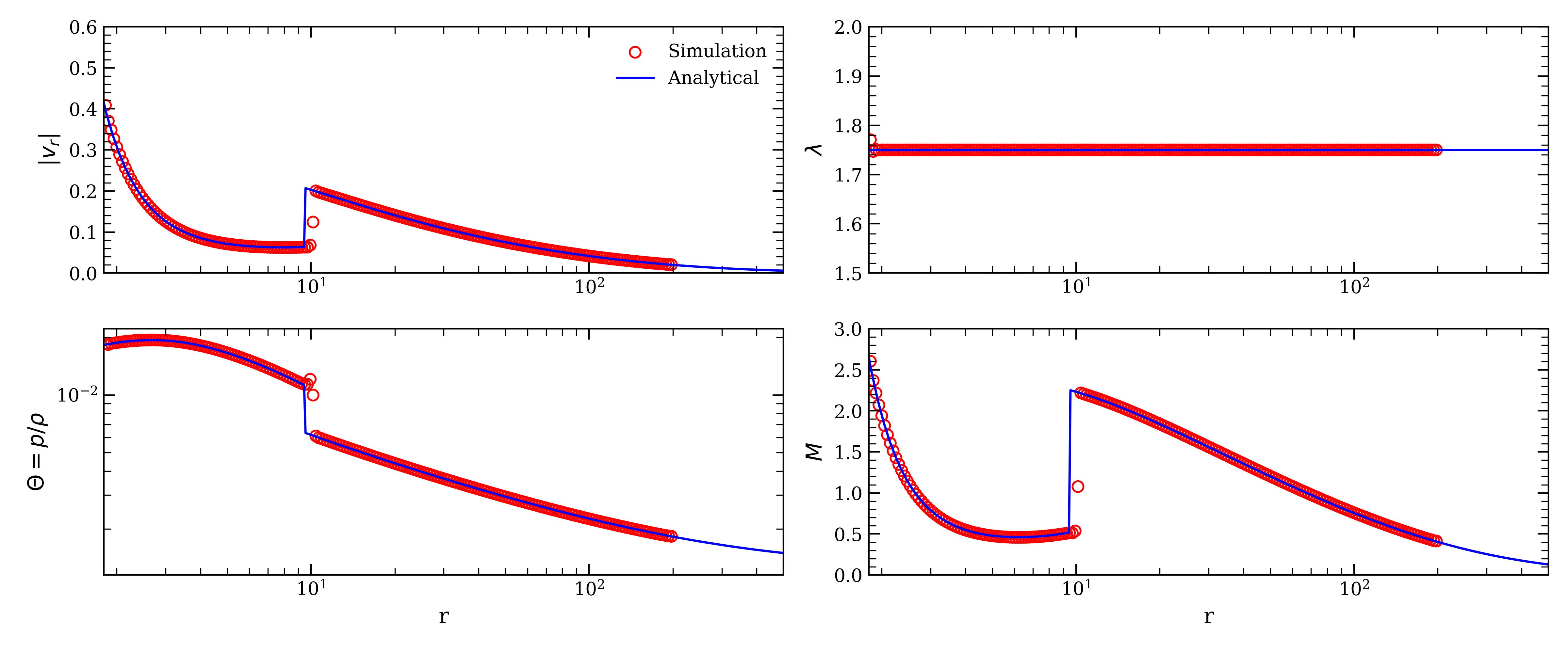}
        \caption{Comparison between one-dimensional simulation and analytical solution.}
        \label{fig:1dver}
        \end{center}
\end{figure}

In numerical simulations, the formation of shocks critically depends on the proper satisfaction of the Rankine–Hugoniot jump conditions. Since shocks inherently generate entropy, their accurate capture and resolution are sensitive to numerical resolution. If the numerical dissipation is large, the code will miss the shock jump. To determine an optimal setup, we performed a one-dimensional simulation to reproduce the analytical solution.
Under the steady-state hydrodynamic approximation, we impose
$\partial/\partial t = \partial/\partial \theta = \partial/\partial \phi = 0$. 
We further assume purely radial inflow, i.e., $v_\theta = 0$, and neglect magnetic fields ($\mathbf{B} = 0$). 
Applying these assumptions to the hydrodynamic conservation equations and performing the necessary algebraic manipulations, the radial velocity gradient can be expressed as
\begin{equation}
\frac{dv_r}{dr}= 
\frac{\displaystyle \frac{2c_s^2}{r} + \frac{\lambda^2}{r^3} - \frac{d\Phi_{PW}}{dr}}
{\displaystyle v_r - \frac{c_s^2}{v_r}},
\label{eq:dvdr}
\end{equation}
where $\lambda = r v_{\phi}$ denotes the specific angular momentum and $c_s$ is the adiabatic sound speed.
For a conical flow geometry, we adopt a constant mass accretion rate given by
\begin{equation}
\dot{M} = -4 \pi \cos{\theta_0}\, r^2 \rho v_r,
\end{equation}
where $\theta_0$ represents the co-latitude of the flow.
Combining the mass conservation equation with the radial momentum and energy equations, one obtains a conserved quantity, commonly referred to as the Bernoulli parameter \citep{1989ApJ...347..365C,2001ApJ...551L..77M,2013MNRAS.430..386K}, which can be written as
\begin{equation}
\mathcal{E} = \frac{v_r^2}{2} 
+ \frac{c_s^2}{\gamma - 1} 
+ \frac{\lambda^2}{2r^2}  
+ \Phi_{PW}(r).
\label{eq:bernoulli}
\end{equation}
This quantity represents the specific total energy of the flow.
For prescribed values of the specific angular momentum $\lambda$ and the Bernoulli parameter $\mathcal{E}$, equation~(\ref{eq:dvdr}) can be integrated to obtain transonic accretion solutions (see, e.g., \citep{1989ApJ...347..365C,1996PhR...266..229C,2013MNRAS.430..386K,2015MNRAS.447.1565S,2024MNRAS.528.3964D}).
As discussed in the Introduction, a global transonic advective accretion solution can admit multiple sonic points. 
Within a specific region of the specific energy ($\mathcal{E}$)--angular momentum ($\lambda$) parameter space, such solutions may also exhibit standing shock waves if Rankine–Hugoniot jump conditions are satisfied. The mathematical formulation for the shock jump conditions of these solutions has been extensively discussed in earlier works.
The region of the $\mathcal{E}$–$\lambda$ parameter space that allows multiple sonic points and shock formation in inviscid accretion flows is illustrated in Fig.~2 of \citet{2024MNRAS.528.3964D}. It is evident from this parameter space that, for a fixed value of $\mathcal{E}$, shock solutions appear only beyond a critical value of $\lambda$. As $\lambda$ is increased while keeping $\mathcal{E}$ fixed within the shock-permitted region, the centrifugal barrier becomes stronger, causing the shock location to move farther away from the black hole.
From this parameter space, we select one representative shock solution with flow parameters $(\mathcal{E}, \lambda) = ( 0.005, 1.75)$, which analytically exhibits a shock transition at a radial location $r = 9.5$, as shown in Fig.~\ref{fig:1dver}.

The results of this comparison are shown in Fig.~\ref{fig:1dver}, where we plot the radial profiles of the absolute radial velocity $v_r$, specific angular momentum $\lambda$, temperature $\Theta = p/\rho$, and Mach number. The simulation results correspond to a resolution of $N_r = 200$. The comparison demonstrates that the shock is captured efficiently within approximately three grid cells. The code also preserves angular momentum very accurately. Although a minor discrepancy in the shock location is present, the solution was found to be stable.
   
\begin{adjustwidth}{-\extralength}{0cm}

\PublishersNote{}
\end{adjustwidth}

\bibliography{biblio}{}

@ARTICLE{2009ApJ...694..492C,
       author = {{Chattopadhyay}, Indranil and {Ryu}, Dongsu},
        title = "{Effects of Fluid Composition on Spherical Flows Around Black Holes}",
      journal = {\apj},
         year = 2009,
        month = mar,
       volume = {694},
       number = {1},
        pages = {492-501},
          doi = {10.1088/0004-637X/694/1/492},
archivePrefix = {arXiv},
       eprint = {0812.2607},
 primaryClass = {astro-ph},
       adsurl = {https://ui.adsabs.harvard.edu/abs/2009ApJ...694..492C}
}

@ARTICLE{2024MNRAS.528.3964D,
       author = {{Debnath}, Sanjit and {Chattopadhyay}, Indranil and {Joshi}, Raj Kishor},
        title = "{Oscillating shocks in the transonic viscous, variable {\ensuremath{\Gamma}} accretion flows around black holes}",
      journal = {\mnras},
         year = 2024,
        month = mar,
       volume = {528},
       number = {3},
        pages = {3964-3980},
          doi = {10.1093/mnras/stae181},
archivePrefix = {arXiv},
       eprint = {2401.07786},
 primaryClass = {astro-ph.HE},
       adsurl = {https://ui.adsabs.harvard.edu/abs/2024MNRAS.528.3964D}
}

@ARTICLE{1999ApJ...521..650B,
       author = {{Balbus}, Steven A. and {Papaloizou}, John C.~B.},
        title = "{On the Dynamical Foundations of {\ensuremath{\alpha}} Disks}",
      journal = {\apj},
         year = 1999,
        month = aug,
       volume = {521},
       number = {2},
        pages = {650-658},
          doi = {10.1086/307594},
archivePrefix = {arXiv},
       eprint = {astro-ph/9903035},
 primaryClass = {astro-ph},
       adsurl = {https://ui.adsabs.harvard.edu/abs/1999ApJ...521..650B}
}

@ARTICLE{2001ApJ...551L..77M,
       author = {{Molteni}, D. and {Gerardi}, G. and {Valenza}, M.~A.},
        title = "{Why Canonical Disks Cannot Produce Advection-dominated Flows}",
      journal = {\apjl},
         year = 2001,
        month = apr,
       volume = {551},
       number = {1},
        pages = {L77-L80},
          doi = {10.1086/319850},
archivePrefix = {arXiv},
       eprint = {astro-ph/0103189},
 primaryClass = {astro-ph},
       adsurl = {https://ui.adsabs.harvard.edu/abs/2001ApJ...551L..77M}
}

@ARTICLE{2008ApJ...677L..93B,
       author = {{Becker}, Peter A. and {Das}, Santabrata and {Le}, Truong},
        title = "{Particle Acceleration and the Formation of Relativistic Outflows in Viscous Accretion Disks with Shocks}",
      journal = {\apjl},
         year = 2008,
        month = apr,
       volume = {677},
       number = {2},
        pages = {L93},
          doi = {10.1086/588137},
archivePrefix = {arXiv},
       eprint = {0907.0872},
 primaryClass = {astro-ph.HE},
       adsurl = {https://ui.adsabs.harvard.edu/abs/2008ApJ...677L..93B}
}

@ARTICLE{2013MNRAS.430..386K,
       author = {{Kumar}, Rajiv and {Chattopadhyay}, Indranil},
        title = "{Estimation of the mass outflow rates from viscous accretion discs}",
      journal = {\mnras},
         year = 2013,
        month = mar,
       volume = {430},
       number = {1},
        pages = {386-402},
          doi = {10.1093/mnras/sts641},
archivePrefix = {arXiv},
       eprint = {1212.4231},
 primaryClass = {astro-ph.HE},
       adsurl = {https://ui.adsabs.harvard.edu/abs/2013MNRAS.430..386K}
}

@ARTICLE{2014MNRAS.443.3444K,
       author = {{Kumar}, Rajiv and {Chattopadhyay}, Indranil},
        title = "{Dissipative advective accretion disc solutions with variable adiabatic index around black holes}",
      journal = {\mnras},
         year = 2014,
        month = oct,
       volume = {443},
       number = {4},
        pages = {3444-3462},
          doi = {10.1093/mnras/stu1389},
archivePrefix = {arXiv},
       eprint = {1407.2130},
 primaryClass = {astro-ph.HE},
       adsurl = {https://ui.adsabs.harvard.edu/abs/2014MNRAS.443.3444K}
}

@ARTICLE{1972A&A....21....1P,
       author = {{Pringle}, J.~E. and {Rees}, M.~J.},
        title = "{Accretion Disc Models for Compact X-Ray Sources}",
      journal = {\aap},
         year = 1972,
        month = oct,
       volume = {21},
        pages = {1},
       adsurl = {https://ui.adsabs.harvard.edu/abs/1972A&A....21....1P}
}

@ARTICLE{1973A&A....24..337S,
       author = {{Shakura}, N.~I. and {Sunyaev}, R.~A.},
        title = "{Black holes in binary systems. Observational appearance.}",
      journal = {\aap},
         year = 1973,
        month = jan,
       volume = {24},
        pages = {337-355},
       adsurl = {https://ui.adsabs.harvard.edu/abs/1973A&A....24..337S}
}

@ARTICLE{1998ApJ...492L..59S,
       author = {{Stella}, Luigi and {Vietri}, Mario},
        title = "{Lense-Thirring Precession and Quasi-periodic Oscillations in Low-Mass X-Ray Binaries}",
      journal = {\apjl},
         year = 1998,
        month = jan,
       volume = {492},
       number = {1},
        pages = {L59-L62},
          doi = {10.1086/311075},
archivePrefix = {arXiv},
       eprint = {astro-ph/9709085},
 primaryClass = {astro-ph},
       adsurl = {https://ui.adsabs.harvard.edu/abs/1998ApJ...492L..59S}
}

@ARTICLE{1999ApJ...524L..63S,
       author = {{Stella}, Luigi and {Vietri}, Mario and {Morsink}, Sharon M.},
        title = "{Correlations in the Quasi-periodic Oscillation Frequencies of Low-Mass X-Ray Binaries and the Relativistic Precession Model}",
      journal = {\apjl},
         year = 1999,
        month = oct,
       volume = {524},
       number = {1},
        pages = {L63-L66},
          doi = {10.1086/312291},
archivePrefix = {arXiv},
       eprint = {astro-ph/9907346},
 primaryClass = {astro-ph},
       adsurl = {https://ui.adsabs.harvard.edu/abs/1999ApJ...524L..63S}
}

@ARTICLE{2009MNRAS.397L.101I,
       author = {{Ingram}, Adam and {Done}, Chris and {Fragile}, P. Chris},
        title = "{Low-frequency quasi-periodic oscillations spectra and Lense-Thirring precession}",
      journal = {\mnras},
         year = 2009,
        month = jul,
       volume = {397},
       number = {1},
        pages = {L101-L105},
          doi = {10.1111/j.1745-3933.2009.00693.x},
archivePrefix = {arXiv},
       eprint = {0901.1238},
 primaryClass = {astro-ph.SR},
       adsurl = {https://ui.adsabs.harvard.edu/abs/2009MNRAS.397L.101I}
}

@ARTICLE{1999A&A...349.1003T,
       author = {{Tagger}, M. and {Pellat}, R.},
        title = "{An accretion-ejection instability in magnetized disks}",
      journal = {\aap},
         year = 1999,
        month = sep,
       volume = {349},
        pages = {1003-1016},
          doi = {10.48550/arXiv.astro-ph/9907267},
archivePrefix = {arXiv},
       eprint = {astro-ph/9907267},
 primaryClass = {astro-ph},
       adsurl = {https://ui.adsabs.harvard.edu/abs/1999A&A...349.1003T}
}

@ARTICLE{2008A&A...489L..41C,
       author = {{Chakrabarti}, S.~K. and {Debnath}, D. and {Nandi}, A. and {Pal}, P.~S.},
        title = "{Evolution of the quasi-periodic oscillation frequency in GRO J1655-40 - Implications for accretion disk dynamics}",
      journal = {\aap},
         year = 2008,
        month = oct,
       volume = {489},
       number = {3},
        pages = {L41-L44},
          doi = {10.1051/0004-6361:200810136},
archivePrefix = {arXiv},
       eprint = {0809.0876},
 primaryClass = {astro-ph},
       adsurl = {https://ui.adsabs.harvard.edu/abs/2008A&A...489L..41C}
}

@ARTICLE{1998MNRAS.299..799L,
       author = {{Lanzafame}, Giuseppe and {Molteni}, D. and {Chakrabarti}, Sandip K.},
        title = "{Smoothed particle hydrodynamic simulations of viscous accretion discs around black holes}",
      journal = {\mnras},
         year = 1998,
        month = sep,
       volume = {299},
       number = {3},
        pages = {799-804},
          doi = {10.1046/j.1365-8711.1998.01816.x},
archivePrefix = {arXiv},
       eprint = {astro-ph/9706248},
 primaryClass = {astro-ph},
       adsurl = {https://ui.adsabs.harvard.edu/abs/1998MNRAS.299..799L}
}

@ARTICLE{2014MNRAS.442..251D,
       author = {{Das}, Santabrata and {Chattopadhyay}, Indranil and {Nandi}, Anuj and {Molteni}, D.},
        title = "{Periodic mass loss from viscous accretion flows around black holes}",
      journal = {\mnras},
         year = 2014,
        month = jul,
       volume = {442},
       number = {1},
        pages = {251-258},
          doi = {10.1093/mnras/stu864},
archivePrefix = {arXiv},
       eprint = {1405.4415},
 primaryClass = {astro-ph.HE},
       adsurl = {https://ui.adsabs.harvard.edu/abs/2014MNRAS.442..251D}
}

@ARTICLE{2011ApJ...728..142L,
       author = {{Lee}, Seong-Jae and {Ryu}, Dongsu and {Chattopadhyay}, Indranil},
        title = "{Quasi-spherical, Time-dependent Viscous Accretion Flow: One-dimensional Results}",
      journal = {\apj},
         year = 2011,
        month = feb,
       volume = {728},
       number = {2},
          eid = {142},
        pages = {142},
          doi = {10.1088/0004-637X/728/2/142},
archivePrefix = {arXiv},
       eprint = {1012.4548},
 primaryClass = {astro-ph.CO},
       adsurl = {https://ui.adsabs.harvard.edu/abs/2011ApJ...728..142L}
}

@ARTICLE{2012MNRAS.421..666G,
       author = {{Giri}, Kinsuk and {Chakrabarti}, Sandip K.},
        title = "{Hydrodynamic simulations of viscous accretion flows around black holes}",
      journal = {\mnras},
         year = 2012,
        month = mar,
       volume = {421},
       number = {1},
        pages = {666-678},
          doi = {10.1111/j.1365-2966.2011.20343.x},
archivePrefix = {arXiv},
       eprint = {1112.1500},
 primaryClass = {astro-ph.HE},
       adsurl = {https://ui.adsabs.harvard.edu/abs/2012MNRAS.421..666G}
}

@ARTICLE{2016ApJ...831...33L,
       author = {{Lee}, Seong-Jae and {Chattopadhyay}, Indranil and {Kumar}, Rajiv and {Hyung}, Siek and {Ryu}, Dongsu},
        title = "{Simulations of Viscous Accretion Flow around Black Holes in a Two-dimensional Cylindrical Geometry}",
      journal = {\apj},
         year = 2016,
        month = nov,
       volume = {831},
       number = {1},
          eid = {33},
        pages = {33},
          doi = {10.3847/0004-637X/831/1/33},
archivePrefix = {arXiv},
       eprint = {1608.03997},
 primaryClass = {astro-ph.HE},
       adsurl = {https://ui.adsabs.harvard.edu/abs/2016ApJ...831...33L}
}

@ARTICLE{1994ApJ...425..161M,
       author = {{Molteni}, Diego and {Lanzafame}, Giuseppe and {Chakrabarti}, Sandip K.},
        title = "{Simulation of Thick Accretion Disks with Standing Shocks by Smoothed Particle Hydrodynamics}",
      journal = {\apj},
         year = 1994,
        month = apr,
       volume = {425},
        pages = {161},
          doi = {10.1086/173972},
archivePrefix = {arXiv},
       eprint = {astro-ph/9310047},
 primaryClass = {astro-ph},
       adsurl = {https://ui.adsabs.harvard.edu/abs/1994ApJ...425..161M}
}

@ARTICLE{2009ApJ...702..649D,
       author = {{Das}, Santabrata and {Becker}, Peter A. and {Le}, Truong},
        title = "{Dynamical Structure of Viscous Accretion Disks with Shocks}",
      journal = {\apj},
         year = 2009,
        month = sep,
       volume = {702},
       number = {1},
        pages = {649-659},
          doi = {10.1088/0004-637X/702/1/649},
archivePrefix = {arXiv},
       eprint = {0907.0875},
 primaryClass = {astro-ph.HE},
       adsurl = {https://ui.adsabs.harvard.edu/abs/2009ApJ...702..649D}
}

@ARTICLE{2017MNRAS.472..542K,
       author = {{Kim}, Jinho and {Garain}, Sudip K. and {Balsara}, Dinshaw S. and {Chakrabarti}, Sandip K.},
        title = "{General relativistic numerical simulation of sub-Keplerian transonic accretion flows on to black holes: Schwarzschild space-time}",
      journal = {\mnras},
         year = 2017,
        month = nov,
       volume = {472},
       number = {1},
        pages = {542-549},
          doi = {10.1093/mnras/stx1986},
archivePrefix = {arXiv},
       eprint = {1707.09856},
 primaryClass = {astro-ph.HE},
       adsurl = {https://ui.adsabs.harvard.edu/abs/2017MNRAS.472..542K}
}

@ARTICLE{2019MNRAS.482.3636K,
       author = {{Kim}, Jinho and {Garain}, Sudip K. and {Chakrabarti}, Sandip K. and {Balsara}, Dinshaw S.},
        title = "{General relativistic numerical simulation of sub-Keplerian transonic accretion flows on to rotating black holes: Kerr space-time}",
      journal = {\mnras},
         year = 2019,
        month = jan,
       volume = {482},
       number = {3},
        pages = {3636-3645},
          doi = {10.1093/mnras/sty2953},
archivePrefix = {arXiv},
       eprint = {1810.12469},
 primaryClass = {astro-ph.HE},
       adsurl = {https://ui.adsabs.harvard.edu/abs/2019MNRAS.482.3636K}
}

@ARTICLE{2023MNRAS.519.4550G,
       author = {{Garain}, Sudip K. and {Kim}, Jinho},
        title = "{Three-dimensional simulations of advective, sub-Keplerian accretion flow on to non-rotating black holes}",
      journal = {\mnras},
         year = 2023,
        month = mar,
       volume = {519},
       number = {3},
        pages = {4550-4563},
          doi = {10.1093/mnras/stac3736},
archivePrefix = {arXiv},
       eprint = {2212.08310},
 primaryClass = {astro-ph.HE},
       adsurl = {https://ui.adsabs.harvard.edu/abs/2023MNRAS.519.4550G}
}

@INPROCEEDINGS{1973blho.conf..343N,
       author = {{Novikov}, I.~D. and {Thorne}, K.~S.},
        title = "{Astrophysics of black holes.}",
    booktitle = {Black Holes (Les Astres Occlus)},
         year = 1973,
        month = jan,
        pages = {343-450},
       adsurl = {https://ui.adsabs.harvard.edu/abs/1973blho.conf..343N}
}

@BOOK{1983JBAA...93R.276S,
       author = {{Shapiro}, Stuart L. and {Teukolsky}, Saul A.},
        title = "{Black holes, white dwarfs and neutron stars. The physics of compact objects}",
         year = 1983,
          doi = {10.1002/9783527617661},
       adsurl = {https://ui.adsabs.harvard.edu/abs/1983bhwd.book.....S}
}

@ARTICLE{2023A&A...678A.141O,
       author = {{Olivares}, H{\'e}ctor R. and {Mo{\'s}cibrodzka}, Monika A. and {Porth}, Oliver},
        title = "{General relativistic hydrodynamic simulations of perturbed transonic accretion}",
      journal = {\aap},
         year = 2023,
        month = oct,
       volume = {678},
          eid = {A141},
        pages = {A141},
          doi = {10.1051/0004-6361/202346010},
archivePrefix = {arXiv},
       eprint = {2301.12020},
 primaryClass = {astro-ph.HE},
       adsurl = {https://ui.adsabs.harvard.edu/abs/2023A&A...678A.141O}
}

@ARTICLE{1987PASJ...39..309F,
       author = {{Fukue}, Jun},
        title = "{Transonic disk accretion revisited}",
      journal = {\pasj},
         year = 1987,
        month = jan,
       volume = {39},
       number = {2},
        pages = {309-327},
       adsurl = {https://ui.adsabs.harvard.edu/abs/1987PASJ...39..309F}
}

@ARTICLE{1989ApJ...347..365C,
       author = {{Chakrabarti}, Sandip K.},
        title = "{Standing Rankine-Hugoniot Shocks in the Hybrid Model Flows of the Black Hole Accretion and Winds}",
      journal = {\apj},
         year = 1989,
        month = dec,
       volume = {347},
        pages = {365},
          doi = {10.1086/168125},
       adsurl = {https://ui.adsabs.harvard.edu/abs/1989ApJ...347..365C}
}

@ARTICLE{1976ApJ...207..962F,
       author = {{Fishbone}, L.~G. and {Moncrief}, V.},
        title = "{Relativistic fluid disks in orbit around Kerr black holes.}",
      journal = {\apj},
         year = 1976,
        month = aug,
       volume = {207},
        pages = {962-976},
          doi = {10.1086/154565},
       adsurl = {https://ui.adsabs.harvard.edu/abs/1976ApJ...207..962F}
}

@ARTICLE{1980A&A....88...23P,
       author = {{Paczy{\'n}sky}, B. and {Wiita}, P.~J.},
        title = "{Thick Accretion Disks and Supercritical Luminosities}",
      journal = {\aap},
         year = 1980,
        month = aug,
       volume = {88},
        pages = {23},
       adsurl = {https://ui.adsabs.harvard.edu/abs/1980A&A....88...23P}
}

@ARTICLE{2019NewAR..8501524I,
       author = {{Ingram}, Adam R. and {Motta}, Sara E.},
        title = "{A review of quasi-periodic oscillations from black hole X-ray binaries: Observation and theory}",
      journal = {\nar},
         year = 2019,
        month = sep,
       volume = {85},
          eid = {101524},
        pages = {101524},
          doi = {10.1016/j.newar.2020.101524},
archivePrefix = {arXiv},
       eprint = {2001.08758},
 primaryClass = {astro-ph.HE},
       adsurl = {https://ui.adsabs.harvard.edu/abs/2019NewAR..8501524I}
}

@ARTICLE{1996ApJ...457..805M,
       author = {{Molteni}, Diego and {Sponholz}, Hanno and {Chakrabarti}, Sandip K.},
        title = "{Resonance Oscillation of Radiative Shock Waves in Accretion Disks around Compact Objects}",
      journal = {\apj},
         year = 1996,
        month = feb,
       volume = {457},
        pages = {805},
          doi = {10.1086/176775},
archivePrefix = {arXiv},
       eprint = {astro-ph/9508022},
 primaryClass = {astro-ph},
       adsurl = {https://ui.adsabs.harvard.edu/abs/1996ApJ...457..805M}
}

@ARTICLE{1996ApJ...470..460M,
       author = {{Molteni}, Diego and {Ryu}, Dongsu and {Chakrabarti}, Sandip K.},
        title = "{Numerical Simulations of Standing Shocks in Accretion Flows around Black Holes: A Comparative Study}",
      journal = {\apj},
         year = 1996,
        month = oct,
       volume = {470},
        pages = {460},
          doi = {10.1086/177877},
archivePrefix = {arXiv},
       eprint = {astro-ph/9605116},
 primaryClass = {astro-ph},
       adsurl = {https://ui.adsabs.harvard.edu/abs/1996ApJ...470..460M}
}

@ARTICLE{1995ApJ...452..364R,
       author = {{Ryu}, Dongsu and {Brown}, Garry L. and {Ostriker}, Jeremiah P. and {Loeb}, Abraham},
        title = "{Stable and Unstable Accretion Flows with Angular Momentum near a Point Mass}",
      journal = {\apj},
         year = 1995,
        month = oct,
       volume = {452},
        pages = {364},
          doi = {10.1086/176308},
archivePrefix = {arXiv},
       eprint = {astro-ph/9504004},
 primaryClass = {astro-ph},
       adsurl = {https://ui.adsabs.harvard.edu/abs/1995ApJ...452..364R}
}

@ARTICLE{2012MNRAS.425.2413O,
       author = {{Okuda}, T. and {Molteni}, D.},
        title = "{Low angular momentum flow model for Sgr A*}",
      journal = {\mnras},
         year = 2012,
        month = oct,
       volume = {425},
       number = {4},
        pages = {2413-2421},
          doi = {10.1111/j.1365-2966.2012.21571.x},
archivePrefix = {arXiv},
       eprint = {1206.5338},
 primaryClass = {astro-ph.HE},
       adsurl = {https://ui.adsabs.harvard.edu/abs/2012MNRAS.425.2413O}
}

@ARTICLE{2022MNRAS.514.5074O,
       author = {{Okuda}, Toru and {Singh}, Chandra B. and {Aktar}, Ramiz},
        title = "{Radiative shock oscillation model for the long-term flares of Sgr A*}",
      journal = {\mnras},
         year = 2022,
        month = aug,
       volume = {514},
       number = {4},
        pages = {5074-5084},
          doi = {10.1093/mnras/stac1630},
archivePrefix = {arXiv},
       eprint = {2206.04919},
 primaryClass = {astro-ph.HE},
       adsurl = {https://ui.adsabs.harvard.edu/abs/2022MNRAS.514.5074O}
}

@ARTICLE{2020ApJ...904...21P,
       author = {{Palit}, Ishika and {Janiuk}, Agnieszka and {Czerny}, Bozena},
        title = "{Clumpy Wind Accretion in Cygnus X-1}",
      journal = {\apj},
         year = 2020,
        month = nov,
       volume = {904},
       number = {1},
          eid = {21},
        pages = {21},
          doi = {10.3847/1538-4357/abba1b},
archivePrefix = {arXiv},
       eprint = {2009.09121},
 primaryClass = {astro-ph.HE},
       adsurl = {https://ui.adsabs.harvard.edu/abs/2020ApJ...904...21P}
}

@ARTICLE{2017MNRAS.472.4327S,
       author = {{Sukov{\'a}}, P. and {Charzy{\'n}ski}, S. and {Janiuk}, A.},
        title = "{Shocks in the relativistic transonic accretion with low angular momentum}",
      journal = {\mnras},
         year = 2017,
        month = dec,
       volume = {472},
       number = {4},
        pages = {4327-4342},
          doi = {10.1093/mnras/stx2254},
archivePrefix = {arXiv},
       eprint = {1709.01824},
 primaryClass = {astro-ph.HE},
       adsurl = {https://ui.adsabs.harvard.edu/abs/2017MNRAS.472.4327S}
}

@ARTICLE{1995ApJ...455..623C,
       author = {{Chakrabarti}, Sandip and {Titarchuk}, Lev G.},
        title = "{Spectral Properties of Accretion Disks around Galactic and Extragalactic Black Holes}",
      journal = {\apj},
         year = 1995,
        month = dec,
       volume = {455},
        pages = {623},
          doi = {10.1086/176610},
archivePrefix = {arXiv},
       eprint = {astro-ph/9510005},
 primaryClass = {astro-ph},
       adsurl = {https://ui.adsabs.harvard.edu/abs/1995ApJ...455..623C}
}

@ARTICLE{2011IJMPD..20.1597C,
       author = {{Chattopadhyay}, Indranil and {Chakrabarti}, Sandip K.},
        title = "{Effects of the Composition on Transonic Properties of Accretion Flows around Black Holes}",
      journal = {International Journal of Modern Physics D},
         year = 2011,
        month = jan,
       volume = {20},
       number = {9},
        pages = {1597-1615},
          doi = {10.1142/S0218271811019487},
archivePrefix = {arXiv},
       eprint = {1309.7900},
 primaryClass = {astro-ph.HE},
       adsurl = {https://ui.adsabs.harvard.edu/abs/2011IJMPD..20.1597C}
}

@ARTICLE{2020A&A...642A.209S,
       author = {{Sarkar}, Shilpa and {Chattopadhyay}, Indranil and {Laurent}, Philippe},
        title = "{Two-temperature solutions and emergent spectra from relativistic accretion discs around black holes}",
      journal = {\aap},
         year = 2020,
        month = oct,
       volume = {642},
          eid = {A209},
        pages = {A209},
          doi = {10.1051/0004-6361/202037520},
archivePrefix = {arXiv},
       eprint = {2007.00919},
 primaryClass = {astro-ph.HE},
       adsurl = {https://ui.adsabs.harvard.edu/abs/2020A&A...642A.209S}
}

@ARTICLE{2023MNRAS.522.3735S,
       author = {{Sarkar}, Shilpa and {Singh}, Kuldeep and {Chattopadhyay}, Indranil and {Laurent}, Philippe},
        title = "{Two-temperature accretion flows around strongly magnetized stars and their spectral analysis}",
      journal = {\mnras},
         year = 2023,
        month = jul,
       volume = {522},
       number = {3},
        pages = {3735-3752},
          doi = {10.1093/mnras/stad1064},
archivePrefix = {arXiv},
       eprint = {2304.03329},
 primaryClass = {astro-ph.HE},
       adsurl = {https://ui.adsabs.harvard.edu/abs/2023MNRAS.522.3735S}
}

@ARTICLE{2024MNRAS.531.1149N,
       author = {{Nandi}, Anuj and {Das}, Santabrata and {Majumder}, Seshadri and {Katoch}, Tilak and {Antia}, H.~M. and {Shah}, Parag},
        title = "{Discovery of evolving low-frequency QPOs in hard X-rays ( 100 keV) observed in black hole Swift J1727.8-1613 with AstroSat}",
      journal = {\mnras},
         year = 2024,
        month = jun,
       volume = {531},
       number = {1},
        pages = {1149-1157},
          doi = {10.1093/mnras/stae1208},
archivePrefix = {arXiv},
       eprint = {2404.17160},
 primaryClass = {astro-ph.HE},
       adsurl = {https://ui.adsabs.harvard.edu/abs/2024MNRAS.531.1149N}
}

@ARTICLE{2022ApJ...936L...5L,
       author = {{Lalakos}, Aretaios and {Gottlieb}, Ore and {Kaaz}, Nicholas and {Chatterjee}, Koushik and {Liska}, Matthew and {Christie}, Ian M. and {Tchekhovskoy}, Alexander and {Zhuravleva}, Irina and {Nokhrina}, Elena},
        title = "{Bridging the Bondi and Event Horizon Scales: 3D GRMHD Simulations Reveal X-shaped Radio Galaxy Morphology}",
      journal = {\apjl},
         year = 2022,
        month = sep,
       volume = {936},
       number = {1},
          eid = {L5},
        pages = {L5},
          doi = {10.3847/2041-8213/ac7bed},
archivePrefix = {arXiv},
       eprint = {2202.08281},
 primaryClass = {astro-ph.HE},
       adsurl = {https://ui.adsabs.harvard.edu/abs/2022ApJ...936L...5L}
}

@ARTICLE{2024ApJ...964...79L,
       author = {{Lalakos}, Aretaios and {Tchekhovskoy}, Alexander and {Bromberg}, Omer and {Gottlieb}, Ore and {Jacquemin-Ide}, Jonatan and {Liska}, Matthew and {Zhang}, Haocheng},
        title = "{Jets with a Twist: The Emergence of FR0 Jets in a 3D GRMHD Simulation of Zero-angular-momentum Black Hole Accretion}",
      journal = {\apj},
         year = 2024,
        month = mar,
       volume = {964},
       number = {1},
          eid = {79},
        pages = {79},
          doi = {10.3847/1538-4357/ad0974},
archivePrefix = {arXiv},
       eprint = {2310.11487},
 primaryClass = {astro-ph.HE},
       adsurl = {https://ui.adsabs.harvard.edu/abs/2024ApJ...964...79L}
}

@ARTICLE{2023ApJ...946L..42K,
       author = {{Kwan}, Tom M. and {Dai}, Lixin and {Tchekhovskoy}, Alexander},
        title = "{The Effects of Gas Angular Momentum on the Formation of Magnetically Arrested Disks and the Launching of Powerful Jets}",
      journal = {\apjl},
         year = 2023,
        month = apr,
       volume = {946},
       number = {2},
          eid = {L42},
        pages = {L42},
          doi = {10.3847/2041-8213/acc334},
archivePrefix = {arXiv},
       eprint = {2211.12726},
 primaryClass = {astro-ph.HE},
       adsurl = {https://ui.adsabs.harvard.edu/abs/2023ApJ...946L..42K}
}

@ARTICLE{2004MNRAS.355.1105F,
       author = {{Fender}, R.~P. and {Belloni}, T.~M. and {Gallo}, E.},
        title = "{Towards a unified model for black hole X-ray binary jets}",
      journal = {\mnras},
         year = 2004,
        month = dec,
       volume = {355},
       number = {4},
        pages = {1105-1118},
          doi = {10.1111/j.1365-2966.2004.08384.x},
archivePrefix = {arXiv},
       eprint = {astro-ph/0409360},
 primaryClass = {astro-ph},
       adsurl = {https://ui.adsabs.harvard.edu/abs/2004MNRAS.355.1105F}
}

@ARTICLE{1952MNRAS.112..195B,
       author = {{Bondi}, H.},
        title = "{On spherically symmetrical accretion}",
      journal = {\mnras},
         year = 1952,
        month = jan,
       volume = {112},
        pages = {195},
          doi = {10.1093/mnras/112.2.195},
       adsurl = {https://ui.adsabs.harvard.edu/abs/1952MNRAS.112..195B}
}

@ARTICLE{2025ApJ...985..135C,
       author = {{Chan}, Ho-Sang and {Dhang}, Prasun and {Dexter}, Jason and {Begelman}, Mitchell C.},
        title = "{The Impact of Plasma Angular Momentum on Magnetically Arrested Flows and Relativistic Jets in Hot Accretion Flows around Black Holes}",
      journal = {\apj},
         year = 2025,
        month = may,
       volume = {985},
       number = {1},
          eid = {135},
        pages = {135},
          doi = {10.3847/1538-4357/adce7f},
archivePrefix = {arXiv},
       eprint = {2504.15489},
 primaryClass = {astro-ph.HE},
       adsurl = {https://ui.adsabs.harvard.edu/abs/2025ApJ...985..135C}
}

@ARTICLE{2015MNRAS.453..147O,
       author = {{Okuda}, Toru and {Das}, Santabrata},
        title = "{Unstable mass-outflows in geometrically thick accretion flows around black holes}",
      journal = {\mnras},
         year = 2015,
        month = oct,
       volume = {453},
       number = {1},
        pages = {147-156},
          doi = {10.1093/mnras/stv1626},
archivePrefix = {arXiv},
       eprint = {1507.04326},
 primaryClass = {astro-ph.HE},
       adsurl = {https://ui.adsabs.harvard.edu/abs/2015MNRAS.453..147O}
}

@ARTICLE{2013MNRAS.430.2836G,
       author = {{Giri}, Kinsuk and {Chakrabarti}, Sandip K.},
        title = "{Hydrodynamic simulation of two-component advective flows around black holes}",
      journal = {\mnras},
         year = 2013,
        month = apr,
       volume = {430},
       number = {4},
        pages = {2836-2843},
          doi = {10.1093/mnras/stt087},
archivePrefix = {arXiv},
       eprint = {1212.6493},
 primaryClass = {astro-ph.HE},
       adsurl = {https://ui.adsabs.harvard.edu/abs/2013MNRAS.430.2836G}
}

@ARTICLE{2012ApJ...758..114G,
       author = {{Garain}, Sudip K. and {Ghosh}, Himadri and {Chakrabarti}, Sandip K.},
        title = "{Effects of Compton Cooling on Outflow in a Two-component Accretion Flow around a Black Hole: Results of a Coupled Monte Carlo Total Variation Diminishing Simulation}",
      journal = {\apj},
         year = 2012,
        month = oct,
       volume = {758},
       number = {2},
          eid = {114},
        pages = {114},
          doi = {10.1088/0004-637X/758/2/114},
archivePrefix = {arXiv},
       eprint = {1210.3515},
 primaryClass = {astro-ph.HE},
       adsurl = {https://ui.adsabs.harvard.edu/abs/2012ApJ...758..114G}
}

@ARTICLE{2015MNRAS.448.3221G,
       author = {{Giri}, Kinsuk and {Garain}, Sudip K. and {Chakrabarti}, Sandip K.},
        title = "{Segregation of a Keplerian disc and sub-Keplerian halo from a transonic flow around a black hole by viscosity and cooling processes}",
      journal = {\mnras},
         year = 2015,
        month = apr,
       volume = {448},
       number = {4},
        pages = {3221-3228},
          doi = {10.1093/mnras/stv223},
archivePrefix = {arXiv},
       eprint = {1502.00455},
 primaryClass = {astro-ph.HE},
       adsurl = {https://ui.adsabs.harvard.edu/abs/2015MNRAS.448.3221G}
}

@ARTICLE{2014MNRAS.437.1329G,
       author = {{Garain}, Sudip K. and {Ghosh}, Himadri and {Chakrabarti}, Sandip K.},
        title = "{Quasi-periodic oscillations in a radiative transonic flow: results of a coupled Monte Carlo-TVD simulation}",
      journal = {\mnras},
         year = 2014,
        month = jan,
       volume = {437},
       number = {2},
        pages = {1329-1336},
          doi = {10.1093/mnras/stt1969},
archivePrefix = {arXiv},
       eprint = {1310.6493},
 primaryClass = {astro-ph.HE},
       adsurl = {https://ui.adsabs.harvard.edu/abs/2014MNRAS.437.1329G}
}

@ARTICLE{2010MNRAS.403..516G,
       author = {{Giri}, Kinsuk and {Chakrabarti}, Sandip K. and {Samanta}, Madan M. and {Ryu}, D.},
        title = "{Hydrodynamic simulations of oscillating shock waves in a sub-Keplerian accretion flow around black holes}",
      journal = {\mnras},
         year = 2010,
        month = mar,
       volume = {403},
       number = {1},
        pages = {516-524},
          doi = {10.1111/j.1365-2966.2009.16147.x},
archivePrefix = {arXiv},
       eprint = {0912.1174},
 primaryClass = {astro-ph.HE},
       adsurl = {https://ui.adsabs.harvard.edu/abs/2010MNRAS.403..516G}
}

@ARTICLE{2025ApJ...990...12M,
       author = {{Mao}, Jirong and {Dihingia}, Indu K. and {Mizuno}, Yosuke and {Nagataki}, Shigehiro},
        title = "{Low-angular-momentum Black Hole Accretion: First General Relativistic Magnetohydrodynamic Evidence of Standing Shocks}",
      journal = {\apj},
         year = 2025,
        month = sep,
       volume = {990},
       number = {1},
          eid = {12},
        pages = {12},
          doi = {10.3847/1538-4357/adf635},
       adsurl = {https://ui.adsabs.harvard.edu/abs/2025ApJ...990...12M}
}

@ARTICLE{2025ApJ...994...48D,
       author = {{Debnath}, Sanjit and {Chattopadhyay}, Indranil and {Joshi}, Raj Kishor and {Laurent}, Philippe and {Tripathi}, Priyesh Kumar and {Khan}, M. Saleem},
        title = "{Dynamical Properties of Oscillating, Viscous, Transonic Accretion Disks around Black Holes}",
      journal = {\apj},
         year = 2025,
        month = nov,
       volume = {994},
       number = {1},
          eid = {48},
        pages = {48},
          doi = {10.3847/1538-4357/ae0ca7},
archivePrefix = {arXiv},
       eprint = {2509.19934},
 primaryClass = {astro-ph.HE},
       adsurl = {https://ui.adsabs.harvard.edu/abs/2025ApJ...994...48D}
}

@ARTICLE{2025arXiv250723187D,
       author = {{Dihingia}, Indu K. and {Uniyal}, Akhil and {Mizuno}, Yosuke},
        title = "{The Fate of Transonic Shocks around Black Holes and their Future Astrophysical Implications}",
      journal = {arXiv e-prints},
         year = 2025,
        month = jul,
          eid = {arXiv:2507.23187},
        pages = {arXiv:2507.23187},
          doi = {10.48550/arXiv.2507.23187},
archivePrefix = {arXiv},
       eprint = {2507.23187},
 primaryClass = {astro-ph.HE},
       adsurl = {https://ui.adsabs.harvard.edu/abs/2025arXiv250723187D}
}

@ARTICLE{1992ApJ...400..595H,
       author = {{Hawley}, John F. and {Balbus}, Steven A.},
        title = "{A Powerful Local Shear Instability in Weakly Magnetized Disks. III. Long-Term Evolution in a Shearing Sheet}",
      journal = {\apj},
         year = 1992,
        month = dec,
       volume = {400},
        pages = {595},
          doi = {10.1086/172021},
       adsurl = {https://ui.adsabs.harvard.edu/abs/1992ApJ...400..595H}
}

@ARTICLE{1991ApJ...376..214B,
       author = {{Balbus}, Steven A. and {Hawley}, John F.},
        title = "{A Powerful Local Shear Instability in Weakly Magnetized Disks. I. Linear Analysis}",
      journal = {\apj},
         year = 1991,
        month = jul,
       volume = {376},
        pages = {214},
          doi = {10.1086/170270},
       adsurl = {https://ui.adsabs.harvard.edu/abs/1991ApJ...376..214B}
}

@ARTICLE{2025arXiv250916796K,
       author = {{Joshi}, Raj Kishor and {Bhake}, Aryan and {Banerjee}, Biswajit and {Vaidya}, Bhargav and {Ruiz}, Milton and {Tsokaros}, Antonios and {Mignone}, Andrea and {Branchesi}, Marica and {Shukla}, Amit and {{\v{C}}emelji{\'c}}, Miljenko},
        title = "{Binary black holes in magnetized AGN disks}",
      journal = {arXiv e-prints},
         year = 2025,
        month = sep,
          eid = {arXiv:2509.16796},
        pages = {arXiv:2509.16796},
          doi = {10.48550/arXiv.2509.16796},
archivePrefix = {arXiv},
       eprint = {2509.16796},
 primaryClass = {astro-ph.HE},
       adsurl = {https://ui.adsabs.harvard.edu/abs/2025arXiv250916796K}
}

@article{Harten_1983,
author = {Harten, Amiram and Lax, Peter D. and Leer, Bram van},
title = {On Upstream Differencing and Godunov-Type Schemes for Hyperbolic Conservation Laws},
journal = {SIAM Review},
volume = {25},
number = {1},
pages = {35-61},
year = {1983},
doi = {10.1137/1025002},
URL = {  https://doi.org/10.1137/1025002 },
eprint = { https://doi.org/10.1137/1025002 }
}

@ARTICLE{cw84,
       author = {{Colella}, P. and {Woodward}, Paul R.},
        title = "{The Piecewise Parabolic Method (PPM) for Gas-Dynamical Simulations}",
      journal = {Journal of Computational Physics},
         year = 1984,
        month = sep,
       volume = {54},
        pages = {174-201},
          doi = {10.1016/0021-9991(84)90143-8},
       adsurl = {https://ui.adsabs.harvard.edu/abs/1984JCoPh..54..174C}
}

@ARTICLE{dedner02,
       author = {{Dedner}, A. and {Kemm}, F. and {Kr{\"o}ner}, D. and {Munz}, C. -D. and {Schnitzer}, T. and {Wesenberg}, M.},
        title = "{Hyperbolic Divergence Cleaning for the MHD Equations}",
      journal = {Journal of Computational Physics},
         year = 2002,
        month = jan,
       volume = {175},
       number = {2},
        pages = {645-673},
          doi = {10.1006/jcph.2001.6961},
       adsurl = {https://ui.adsabs.harvard.edu/abs/2002JCoPh.175..645D}
}

@ARTICLE{mitra24,
       author = {{Mitra}, Samik and {Das}, Santabrata},
        title = "{Low-angular-momentum General Relativistic Magnetohydrodynamic Accretion Flows around Rotating Black Holes with Shocks}",
      journal = {\apj},
         year = 2024,
        month = aug,
       volume = {971},
       number = {1},
          eid = {28},
        pages = {28},
          doi = {10.3847/1538-4357/ad55cb},
archivePrefix = {arXiv},
       eprint = {2405.16326},
 primaryClass = {astro-ph.HE},
       adsurl = {https://ui.adsabs.harvard.edu/abs/2024ApJ...971...28M}
}

@ARTICLE{2020MNRAS.492.1855M,
       author = {{Mishra}, Bhupendra and {Begelman}, Mitchell C. and {Armitage}, Philip J. and {Simon}, Jacob B.},
        title = "{Strongly magnetized accretion discs: structure and accretion from global magnetohydrodynamic simulations}",
      journal = {\mnras},
         year = 2020,
        month = feb,
       volume = {492},
       number = {2},
        pages = {1855-1868},
          doi = {10.1093/mnras/stz3572},
archivePrefix = {arXiv},
       eprint = {1907.08995},
 primaryClass = {astro-ph.HE},
       adsurl = {https://ui.adsabs.harvard.edu/abs/2020MNRAS.492.1855M}
}

@ARTICLE{2024MNRAS.527.1745A,
       author = {{Aktar}, Ramiz and {Pan}, Kuo-Chuan and {Okuda}, Toru},
        title = "{Evolution of MHD torus and mass outflow around spinning AGNs}",
      journal = {\mnras},
         year = 2024,
        month = jan,
       volume = {527},
       number = {2},
        pages = {1745-1759},
          doi = {10.1093/mnras/stad3287},
archivePrefix = {arXiv},
       eprint = {2310.15501},
 primaryClass = {astro-ph.HE},
       adsurl = {https://ui.adsabs.harvard.edu/abs/2024MNRAS.527.1745A}
}

@ARTICLE{2024ApJ...972...18A,
       author = {{Aktar}, Ramiz and {Pan}, Kuo-Chuan and {Okuda}, Toru},
        title = "{Radiation RMHD Accretion Flows around Spinning AGNs: A Comparative Study of MAD and SANE State}",
      journal = {\apj},
         year = 2024,
        month = sep,
       volume = {972},
       number = {1},
          eid = {18},
        pages = {18},
          doi = {10.3847/1538-4357/ad5a8a},
archivePrefix = {arXiv},
       eprint = {2406.10496},
 primaryClass = {astro-ph.HE},
       adsurl = {https://ui.adsabs.harvard.edu/abs/2024ApJ...972...18A}
}

@ARTICLE{pluto07,
       author = {{Mignone}, A. and {Bodo}, G. and {Massaglia}, S. and {Matsakos}, T. and {Tesileanu}, O. and {Zanni}, C. and {Ferrari}, A.},
        title = "{PLUTO: A Numerical Code for Computational Astrophysics}",
      journal = {\apjs},
         year = 2007,
        month = may,
       volume = {170},
       number = {1},
        pages = {228-242},
          doi = {10.1086/513316},
archivePrefix = {arXiv},
       eprint = {astro-ph/0701854},
 primaryClass = {astro-ph},
       adsurl = {https://ui.adsabs.harvard.edu/abs/2007ApJS..170..228M}
}

@ARTICLE{rc06,
       author = {{Ryu}, Dongsu and {Chattopadhyay}, Indranil and {Choi}, Eunwoo},
        title = "{Equation of State in Numerical Relativistic Hydrodynamics}",
      journal = {\apjs},
         year = 2006,
        month = sep,
       volume = {166},
       number = {1},
        pages = {410-420},
          doi = {10.1086/505937},
archivePrefix = {arXiv},
       eprint = {astro-ph/0605550},
 primaryClass = {astro-ph},
       adsurl = {https://ui.adsabs.harvard.edu/abs/2006ApJS..166..410R}
}

@ARTICLE{m71,
       author = {{Mathews}, William G.},
        title = "{The Hydromagnetic Free Expansion of a Relativistic Gas}",
      journal = {\apj},
         year = 1971,
        month = apr,
       volume = {165},
        pages = {147},
          doi = {10.1086/150883},
       adsurl = {https://ui.adsabs.harvard.edu/abs/1971ApJ...165..147M}
}

@ARTICLE{2004ApJ...605..321S,
       author = {{Sano}, Takayoshi and {Inutsuka}, Shu-ichiro and {Turner}, Neal J. and {Stone}, James M.},
        title = "{Angular Momentum Transport by Magnetohydrodynamic Turbulence in Accretion Disks: Gas Pressure Dependence of the Saturation Level of the Magnetorotational Instability}",
      journal = {\apj},
         year = 2004,
        month = apr,
       volume = {605},
       number = {1},
        pages = {321-339},
          doi = {10.1086/382184},
archivePrefix = {arXiv},
       eprint = {astro-ph/0312480},
 primaryClass = {astro-ph},
       adsurl = {https://ui.adsabs.harvard.edu/abs/2004ApJ...605..321S}
}

@ARTICLE{2025ApJ...978..148G,
       author = {{Galishnikova}, Alisa and {Philippov}, Alexander and {Quataert}, Eliot and {Chatterjee}, Koushik and {Liska}, Matthew},
        title = "{Strongly Magnetized Accretion with Low Angular Momentum Produces a Weak Jet}",
      journal = {\apj},
         year = 2025,
        month = jan,
       volume = {978},
       number = {2},
          eid = {148},
        pages = {148},
          doi = {10.3847/1538-4357/ad9926},
archivePrefix = {arXiv},
       eprint = {2409.11486},
 primaryClass = {astro-ph.HE},
       adsurl = {https://ui.adsabs.harvard.edu/abs/2025ApJ...978..148G}
}

@ARTICLE{1984ApJS...55..211H,
       author = {{Hawley}, J.~F. and {Smarr}, L.~L. and {Wilson}, J.~R.},
        title = "{A numerical study of nonspherical black hole accretion. II - Finite differencing and code calibration}",
      journal = {\apjs},
         year = 1984,
        month = jun,
       volume = {55},
        pages = {211-246},
          doi = {10.1086/190953},
       adsurl = {https://ui.adsabs.harvard.edu/abs/1984ApJS...55..211H}
}

@ARTICLE{2003ApJ...589..444G,
       author = {{Gammie}, Charles F. and {McKinney}, Jonathan C. and {T{\'o}th}, G{\'a}bor},
        title = "{HARM: A Numerical Scheme for General Relativistic Magnetohydrodynamics}",
      journal = {\apj},
         year = 2003,
        month = may,
       volume = {589},
       number = {1},
        pages = {444-457},
          doi = {10.1086/374594},
archivePrefix = {arXiv},
       eprint = {astro-ph/0301509},
 primaryClass = {astro-ph},
       adsurl = {https://ui.adsabs.harvard.edu/abs/2003ApJ...589..444G}
}

@ARTICLE{2009ApJ...703L.142D,
       author = {{Dexter}, Jason and {Agol}, Eric and {Fragile}, P. Chris},
        title = "{Millimeter Flares and VLBI Visibilities from Relativistic Simulations of Magnetized Accretion Onto the Galactic Center Black Hole}",
      journal = {\apjl},
         year = 2009,
        month = oct,
       volume = {703},
       number = {2},
        pages = {L142-L146},
          doi = {10.1088/0004-637X/703/2/L142},
archivePrefix = {arXiv},
       eprint = {0909.0267},
 primaryClass = {astro-ph.HE},
       adsurl = {https://ui.adsabs.harvard.edu/abs/2009ApJ...703L.142D}
}

@ARTICLE{2016A&A...586A..38M,
       author = {{Mo{\'s}cibrodzka}, Monika and {Falcke}, Heino and {Shiokawa}, Hotaka},
        title = "{General relativistic magnetohydrodynamical simulations of the jet in M 87}",
      journal = {\aap},
         year = 2016,
        month = feb,
       volume = {586},
          eid = {A38},
        pages = {A38},
          doi = {10.1051/0004-6361/201526630},
archivePrefix = {arXiv},
       eprint = {1510.07243},
 primaryClass = {astro-ph.HE},
       adsurl = {https://ui.adsabs.harvard.edu/abs/2016A&A...586A..38M}
}

@ARTICLE{2011MNRAS.418L..79T,
       author = {{Tchekhovskoy}, Alexander and {Narayan}, Ramesh and {McKinney}, Jonathan C.},
        title = "{Efficient generation of jets from magnetically arrested accretion on a rapidly spinning black hole}",
      journal = {\mnras},
         year = 2011,
        month = nov,
       volume = {418},
       number = {1},
        pages = {L79-L83},
          doi = {10.1111/j.1745-3933.2011.01147.x},
archivePrefix = {arXiv},
       eprint = {1108.0412},
 primaryClass = {astro-ph.HE},
       adsurl = {https://ui.adsabs.harvard.edu/abs/2011MNRAS.418L..79T}
}

@ARTICLE{2014MNRAS.441.3177M,
       author = {{McKinney}, Jonathan C. and {Tchekhovskoy}, Alexander and {Sadowski}, Aleksander and {Narayan}, Ramesh},
        title = "{Three-dimensional general relativistic radiation magnetohydrodynamical simulation of super-Eddington accretion, using a new code HARMRAD with M1 closure}",
      journal = {\mnras},
         year = 2014,
        month = jul,
       volume = {441},
       number = {4},
        pages = {3177-3208},
          doi = {10.1093/mnras/stu762},
archivePrefix = {arXiv},
       eprint = {1312.6127},
 primaryClass = {astro-ph.CO},
       adsurl = {https://ui.adsabs.harvard.edu/abs/2014MNRAS.441.3177M}
}

@ARTICLE{2012ApJS..201....9F,
       author = {{Fragile}, P. Chris and {Gillespie}, Anna and {Monahan}, Timothy and {Rodriguez}, Marco and {Anninos}, Peter},
        title = "{Numerical Simulations of Optically Thick Accretion onto a Black Hole. I. Spherical Case}",
      journal = {\apjs},
         year = 2012,
        month = aug,
       volume = {201},
       number = {2},
          eid = {9},
        pages = {9},
          doi = {10.1088/0067-0049/201/2/9},
archivePrefix = {arXiv},
       eprint = {1204.5538},
 primaryClass = {astro-ph.IM},
       adsurl = {https://ui.adsabs.harvard.edu/abs/2012ApJS..201....9F}
}

@ARTICLE{2007ApJ...668..417F,
       author = {{Fragile}, P. Chris and {Blaes}, Omer M. and {Anninos}, Peter and {Salmonson}, Jay D.},
        title = "{Global General Relativistic Magnetohydrodynamic Simulation of a Tilted Black Hole Accretion Disk}",
      journal = {\apj},
         year = 2007,
        month = oct,
       volume = {668},
       number = {1},
        pages = {417-429},
          doi = {10.1086/521092},
archivePrefix = {arXiv},
       eprint = {0706.4303},
 primaryClass = {astro-ph},
       adsurl = {https://ui.adsabs.harvard.edu/abs/2007ApJ...668..417F}
}

@ARTICLE{2014MNRAS.439..503S,
       author = {{S{\k{a}}dowski}, Aleksander and {Narayan}, Ramesh and {McKinney}, Jonathan C. and {Tchekhovskoy}, Alexander},
        title = "{Numerical simulations of super-critical black hole accretion flows in general relativity}",
      journal = {\mnras},
         year = 2014,
        month = mar,
       volume = {439},
       number = {1},
        pages = {503-520},
          doi = {10.1093/mnras/stt2479},
archivePrefix = {arXiv},
       eprint = {1311.5900},
 primaryClass = {astro-ph.HE},
       adsurl = {https://ui.adsabs.harvard.edu/abs/2014MNRAS.439..503S}
}

@ARTICLE{2015MNRAS.447...49S,
       author = {{S{\k{a}}dowski}, Aleksander and {Narayan}, Ramesh and {Tchekhovskoy}, Alexander and {Abarca}, David and {Zhu}, Yucong and {McKinney}, Jonathan C.},
        title = "{Global simulations of axisymmetric radiative black hole accretion discs in general relativity with a mean-field magnetic dynamo}",
      journal = {\mnras},
         year = 2015,
        month = feb,
       volume = {447},
       number = {1},
        pages = {49-71},
          doi = {10.1093/mnras/stu2387},
archivePrefix = {arXiv},
       eprint = {1407.4421},
 primaryClass = {astro-ph.HE},
       adsurl = {https://ui.adsabs.harvard.edu/abs/2015MNRAS.447...49S}
}

@ARTICLE{2018MNRAS.474L..81L,
       author = {{Liska}, M. and {Hesp}, C. and {Tchekhovskoy}, A. and {Ingram}, A. and {van der Klis}, M. and {Markoff}, S.},
        title = "{Formation of precessing jets by tilted black hole discs in 3D general relativistic MHD simulations}",
      journal = {\mnras},
         year = 2018,
        month = feb,
       volume = {474},
       number = {1},
        pages = {L81-L85},
          doi = {10.1093/mnrasl/slx174},
archivePrefix = {arXiv},
       eprint = {1707.06619},
 primaryClass = {astro-ph.HE},
       adsurl = {https://ui.adsabs.harvard.edu/abs/2018MNRAS.474L..81L}
}

@ARTICLE{2025ApJ...980..203D,
       author = {{Dhang}, Prasun and {Dexter}, Jason and {Begelman}, Mitchell C.},
        title = "{Energy Extraction from a Black Hole by a Strongly Magnetized Thin Accretion Disk}",
      journal = {\apj},
         year = 2025,
        month = feb,
       volume = {980},
       number = {2},
          eid = {203},
        pages = {203},
          doi = {10.3847/1538-4357/ada76e},
archivePrefix = {arXiv},
       eprint = {2411.02515},
 primaryClass = {astro-ph.HE},
       adsurl = {https://ui.adsabs.harvard.edu/abs/2025ApJ...980..203D}
}

@ARTICLE{2019ApJS..243...26P,
       author = {{Porth}, Oliver and {Chatterjee}, Koushik and {Narayan}, Ramesh and {Gammie}, Charles F. and {Mizuno}, Yosuke and {Anninos}, Peter and {Baker}, John G. and {Bugli}, Matteo and {Chan}, Chi-kwan and {Davelaar}, Jordy and {Del Zanna}, Luca and {Etienne}, Zachariah B. and {Fragile}, P. Chris and {Kelly}, Bernard J. and {Liska}, Matthew and {Markoff}, Sera and {McKinney}, Jonathan C. and {Mishra}, Bhupendra and {Noble}, Scott C. and {Olivares}, H{\'e}ctor and {Prather}, Ben and {Rezzolla}, Luciano and {Ryan}, Benjamin R. and {Stone}, James M. and {Tomei}, Niccol{\`o} and {White}, Christopher J. and {Younsi}, Ziri and {Akiyama}, Kazunori and {Alberdi}, Antxon and {Alef}, Walter and {Asada}, Keiichi and {Azulay}, Rebecca and {Baczko}, Anne-Kathrin and {Ball}, David and {Balokovi{\'c}}, Mislav and {Barrett}, John and {Bintley}, Dan and {Blackburn}, Lindy and {Boland}, Wilfred and {Bouman}, Katherine L. and {Bower}, Geoffrey C. and {Bremer}, Michael and {Brinkerink}, Christiaan D. and {Brissenden}, Roger and {Britzen}, Silke and {Broderick}, Avery E. and {Broguiere}, Dominique and {Bronzwaer}, Thomas and {Byun}, Do-Young and {Carlstrom}, John E. and {Chael}, Andrew and {Chatterjee}, Shami and {Chen}, Ming-Tang and {Chen}, Yongjun and {Cho}, Ilje and {Christian}, Pierre and {Conway}, John E. and {Cordes}, James M. and {Geoffrey} and {Crew}, B. and {Cui}, Yuzhu and {De Laurentis}, Mariafelicia and {Deane}, Roger and {Dempsey}, Jessica and {Desvignes}, Gregory and {Doeleman}, Sheperd S. and {Eatough}, Ralph P. and {Falcke}, Heino and {Fish}, Vincent L. and {Fomalont}, Ed and {Fraga-Encinas}, Raquel and {Freeman}, Bill and {Friberg}, Per and {Fromm}, Christian M. and {G{\'o}mez}, Jos{\'e} L. and {Galison}, Peter and {Garc{\'\i}a}, Roberto and {Gentaz}, Olivier and {Georgiev}, Boris and {Goddi}, Ciriaco and {Gold}, Roman and {Gu}, Minfeng and {Gurwell}, Mark and {Hada}, Kazuhiro and {Hecht}, Michael H. and {Hesper}, Ronald and {Ho}, Luis C. and {Ho}, Paul and {Honma}, Mareki and {Huang}, Chih-Wei L. and {Huang}, Lei and {Hughes}, David H. and {Ikeda}, Shiro and {Inoue}, Makoto and {Issaoun}, Sara and {James}, David J. and {Jannuzi}, Buell T. and {Janssen}, Michael and {Jeter}, Britton and {Jiang}, Wu and {Johnson}, Michael D. and {Jorstad}, Svetlana and {Jung}, Taehyun and {Karami}, Mansour and {Karuppusamy}, Ramesh and {Kawashima}, Tomohisa and {Keating}, Garrett K. and {Kettenis}, Mark and {Kim}, Jae-Young and {Kim}, Junhan and {Kim}, Jongsoo and {Kino}, Motoki and {Koay}, Jun Yi and {Patrick} and {Koch}, M. and {Koyama}, Shoko and {Kramer}, Michael and {Kramer}, Carsten and {Krichbaum}, Thomas P. and {Kuo}, Cheng-Yu and {Lauer}, Tod R. and {Lee}, Sang-Sung and {Li}, Yan-Rong and {Li}, Zhiyuan and {Lindqvist}, Michael and {Liu}, Kuo and {Liuzzo}, Elisabetta and {Lo}, Wen-Ping and {Lobanov}, Andrei P. and {Loinard}, Laurent and {Lonsdale}, Colin and {Lu}, Ru-Sen and {MacDonald}, Nicholas R. and {Mao}, Jirong and {Marrone}, Daniel P. and {Marscher}, Alan P. and {Mart{\'\i}-Vidal}, Iv{\'a}n and {Matsushita}, Satoki and {Matthews}, Lynn D. and {Medeiros}, Lia and {Menten}, Karl M. and {Mizuno}, Izumi and {Moran}, James M. and {Moriyama}, Kotaro and {Moscibrodzka}, Monika and {M{\"u}ller}, Cornelia and {Nagai}, Hiroshi and {Nagar}, Neil M. and {Nakamura}, Masanori and {Narayanan}, Gopal and {Natarajan}, Iniyan and {Neri}, Roberto and {Ni}, Chunchong and {Noutsos}, Aristeidis and {Okino}, Hiroki and {Oyama}, Tomoaki and {{\"O}zel}, Feryal and {Palumbo}, Daniel C.~M. and {Patel}, Nimesh and {Pen}, Ue-Li and {Pesce}, Dominic W. and {Pi{\'e}tu}, Vincent and {Plambeck}, Richard and {PopStefanija}, Aleksandar and {Preciado-L{\'o}pez}, Jorge A. and {Psaltis}, Dimitrios and {Pu}, Hung-Yi and {Ramakrishnan}, Venkatessh and {Rao}, Ramprasad and {Rawlings}, Mark G. and {Raymond}, Alexander W. and {Ripperda}, Bart and {Roelofs}, Freek and {Rogers}, Alan and {Ros}, Eduardo and {Rose}, Mel and {Roshanineshat}, Arash and {Rottmann}, Helge and {Roy}, Alan L. and {Ruszczyk}, Chet and {Rygl}, Kazi L.~J. and {S{\'a}nchez}, Salvador and {S{\'a}nchez-Arguelles}, David and {Sasada}, Mahito and {Savolainen}, Tuomas and {Schloerb}, F. Peter and {Schuster}, Karl-Friedrich and {Shao}, Lijing and {Shen}, Zhiqiang and {Small}, Des and {Sohn}, Bong Won and {SooHoo}, Jason and {Tazaki}, Fumie and {Tiede}, Paul and {Tilanus}, Remo P.~J. and {Titus}, Michael and {Toma}, Kenji and {Torne}, Pablo and {Trent}, Tyler and {Trippe}, Sascha},
        title = "{The Event Horizon General Relativistic Magnetohydrodynamic Code Comparison Project}",
      journal = {\apjs},
         year = 2019,
        month = aug,
       volume = {243},
       number = {2},
          eid = {26},
        pages = {26},
          doi = {10.3847/1538-4365/ab29fd},
archivePrefix = {arXiv},
       eprint = {1904.04923},
 primaryClass = {astro-ph.HE},
       adsurl = {https://ui.adsabs.harvard.edu/abs/2019ApJS..243...26P}
}

@ARTICLE{1996PhR...266..229C,
       author = {{Chakrabarti}, S.~K.},
        title = "{Accretion processes on a black hole.}",
      journal = {\physrep},
         year = 1996,
        month = jan,
       volume = {266},
        pages = {229-390},
          doi = {10.1016/0370-1573(95)00057-7},
archivePrefix = {arXiv},
       eprint = {astro-ph/9605015},
 primaryClass = {astro-ph},
       adsurl = {https://ui.adsabs.harvard.edu/abs/1996PhR...266..229C}
}

@ARTICLE{2015MNRAS.447.1565S,
       author = {{Sukov{\'a}}, P. and {Janiuk}, A.},
        title = "{Oscillating shocks in the low angular momentum flows as a source of variability of accreting black holes}",
      journal = {\mnras},
         year = 2015,
        month = feb,
       volume = {447},
       number = {2},
        pages = {1565-1579},
          doi = {10.1093/mnras/stu2544},
archivePrefix = {arXiv},
       eprint = {1411.7836},
 primaryClass = {astro-ph.HE},
       adsurl = {https://ui.adsabs.harvard.edu/abs/2015MNRAS.447.1565S}
}

@ARTICLE{2012MNRAS.426.3241N,
       author = {{Narayan}, Ramesh and {S{\k{a}}dowski}, Aleksander and {Penna}, Robert F. and {Kulkarni}, Akshay K.},
        title = "{GRMHD simulations of magnetized advection-dominated accretion on a non-spinning black hole: role of outflows}",
      journal = {\mnras},
         year = 2012,
        month = nov,
       volume = {426},
       number = {4},
        pages = {3241-3259},
          doi = {10.1111/j.1365-2966.2012.22002.x},
archivePrefix = {arXiv},
       eprint = {1206.1213},
 primaryClass = {astro-ph.HE},
       adsurl = {https://ui.adsabs.harvard.edu/abs/2012MNRAS.426.3241N}
}

@ARTICLE{2025arXiv251203443Z,
       author = {{Zhou}, Hongzhe and {Mizuno}, Yosuke and {Zhu}, Zhenyu},
        title = "{Subgrid Mean-field Dynamo Model with Dynamical Quenching in General Relativistic Magnetohydrodynamic Simulations}",
      journal = {arXiv e-prints},
         year = 2025,
        month = dec,
          eid = {arXiv:2512.03443},
        pages = {arXiv:2512.03443},
          doi = {10.48550/arXiv.2512.03443},
archivePrefix = {arXiv},
       eprint = {2512.03443},
 primaryClass = {astro-ph.HE},
       adsurl = {https://ui.adsabs.harvard.edu/abs/2025arXiv251203443Z}
}

@ARTICLE{2025ApJ...995..112J,
       author = {{Jiang}, Hong-Xuan and {Mizuno}, Yosuke and {Lai}, Dong and {Dihingia}, Indu K. and {Fromm}, Christian M.},
        title = "{Magnetically Driven Retrograde Precession in Misaligned Black Hole Accretion Flows}",
      journal = {\apj},
         year = 2025,
        month = dec,
       volume = {995},
       number = {1},
          eid = {112},
        pages = {112},
          doi = {10.3847/1538-4357/ae19ee},
archivePrefix = {arXiv},
       eprint = {2507.13680},
 primaryClass = {astro-ph.HE},
       adsurl = {https://ui.adsabs.harvard.edu/abs/2025ApJ...995..112J}
}

@ARTICLE{2025PhRvD.112f3013C,
       author = {{Chatterjee}, Koushik and {Kaaz}, Nicholas and {Liska}, Matthew and {Tchekhovskoy}, Alexander and {Markoff}, Sera},
        title = "{Misaligned magnetized accretion flows onto spinning black holes: Magneto-spin alignment, outflow power, and intermittent jets}",
      journal = {\prd},
         year = 2025,
        month = sep,
       volume = {112},
       number = {6},
          eid = {063013},
        pages = {063013},
          doi = {10.1103/hgj9-v4fk},
archivePrefix = {arXiv},
       eprint = {2311.00432},
 primaryClass = {astro-ph.HE},
       adsurl = {https://ui.adsabs.harvard.edu/abs/2025PhRvD.112f3013C}
}

\end{document}